# The Impact of Supermassive Black Holes on Exoplanet Habitability. I. Spanning the Natural Mass Range

Jourdan Waas[1], Eric S. Perlman[1], Manasvi Lingam[1], Emily Lohmann[1], Jackson Kernan[1], Francesco Tombesi[2,3,4], Amedeo Balbi[2], and Alessandra Ambrifi[5,6]

[1] Department of Aerospace, Physics and Space Sciences, Florida Institute of Technology, Melbourne, FL 32901, USA; waasj2019@my.fit.edu
[2] Physics Department, Tor Vergata University of Rome, Via della Ricerca Scientifica 1, 00133 Rome, Italy
[3] INAF—Astronomical Observatory of Rome, Via Frascati 33, 00040 Monte Porzio Catone, Italy
[4] INFN—Rome Tor Vergata, Via della Ricerca Scientifica 1, 00133 Rome, Italy
[5] Instituto de Astrofísica de Canarias, E-38205 La Laguna, Tenerife, Spain
[6] Departamento de Astrofísica, Universidad de La Laguna, E-38206 La Laguna, Tenerife, Spain
*Received 2025 November 13; revised 2026 April 9; accepted 2026 April 9; published 2026 May 12*

## Abstract

While the influence of supermassive black hole (SMBH) activity on habitability has garnered attention, the specific effects of active galactic nucleus (AGN) winds, particularly ultrafast outflows (UFOs), on planetary atmospheres remain largely unexplored. This study aims to fill this gap by investigating the relationship between SMBH mass at the galactic center and exoplanetary habitability, given that SMBH masses are empirically confirmed to span approximately 5 orders of magnitude in galaxies. Through simplified models, we account for various results involving the relationships between the distance from the planet to the central SMBH and the mass of the SMBH. Specifically, we show that increased SMBH mass leads to higher atmospheric heating and elevated temperatures, greater molecular thermal velocities, and enhanced mass loss, all of which diminish with distance from the galactic center. Energy-driven winds consistently have a stronger impact than momentum-driven ones. Crucially, ozone depletion is shown to rise with SMBH mass and decrease with distance from the galactic center, with nearly complete ozone loss ($\sim$100%) occurring across galactic scales for SMBH masses $\geqslant 10^8\,M_\odot$ in the energy-driven case. This study emphasizes that SMBH growth over cosmic time may have produced markedly different impacts on galactic habitability, depending on both the mass of the central black hole and the location of planetary systems within their host galaxies.

## 1. Introduction

In recent years, considerable attention has been devoted to understanding the role of high-energy astrophysical events in shaping the habitability of galaxies. Supernovae have long captured the attention of researchers due to their profound implications for planetary habitability (N. Gehrels et al. 2003; M. Beech 2011; A. Hanslmeier 2017; A. L. Melott et al. 2017; I. R. Brunton et al. 2023; B. C. Thomas & A. M. Yelland 2023), both due to the enhanced luminosity of the star as well as the shock. There has also been a growing interest surrounding the impact of supermassive black hole (SMBH) activity, e.g., active galactic nuclei (AGN) and associated phenomena (A. Balbi & F. Tombesi 2017; H. Chen et al. 2018; J. C. Forbes & A. Loeb 2018; P. Amaro-Seoane & X. Chen 2019; M. Lingam 2019; M. Lingam et al. 2019; A. M. Wisłocka et al. 2019; C. Liu et al. 2020; A. Ambrifi et al. 2022). A clear understanding of the myriad roles of SMBH activity on galactic habitability (N. Prantzos 2008; M. Scherf et al. 2024; M. Lingam & A. Balbi 2024; C. H. Lineweaver 2025) would help pave the way for gauging the prospects for extraterrestrial habitability and life in the Universe.

However, amidst this growing interest, a significant aspect of the interaction between AGN winds and planetary atmospheres remains insufficiently explored. In particular, one important effect that needs exploring is that from different black hole (BH) masses, as more massive SMBHs yield more luminous AGN. Furthermore, all previous works in this field, barring the paper by A. M. Wisłocka et al. (2019), have concentrated on possible AGN outbursts of the Milky Way's SMBH, Sagittarius A$^*$ (Sgr A$^*$), which has a mass of $4.31 \pm 0.38 \times 10^6\,M_\odot$ (A. M. Ghez et al. 2008; S. Gillessen et al. 2009; A. Boehle et al. 2016; Event Horizon Telescope Collaboration et al. 2022). Although this serves as a valuable local benchmark, it does not capture the full range of AGN environments present across different galactic systems.

Central BHs, found in nearly all galaxies, exhibit masses spanning several orders of magnitude, from intermediate-mass BHs ($10^4$–$10^6\,M_\odot$) to the most massive SMBHs reaching up to ($10^{10}\,M_\odot$; A. D. Dolgov 2020; M. Vestergaard & K. Gültekin 2023). This vast diversity in mass directly influences both the luminosity output and the kinetic power of AGN-driven winds, thereby modifying the degree to which planetary atmospheres may be altered in ways that affect their capacity to support surface life.

To understand the impact of the SMBH in other galaxies, as well as the complex interplay between galactic-scale phenomena and the potential for life-sustaining conditions beyond our solar system, it is necessary to investigate the impact of the SMBH mass on the habitability of exoplanets across galactic scales. This is one of the primary objectives of the paper.

### 1.1. Active Galactic Nucleus Impacts on Habitability

There has been a recent influx of interest pertaining to the impact of the central SMBH on galactic habitability, specifically regarding aspects of AGN (A. Balbi & F. Tombesi 2017; H. Chen et al. 2018; J. C. Forbes & A. Loeb 2018; P. Amaro-Seoane & X. Chen 2019; M. Lingam 2019; M. Lingam et al. 2019; A. M. Wisłocka et al. 2019; C. Liu et al. 2020; A. Ambrifi et al. 2022; W. Ishibashi 2024; K. I. Sippy et al. 2025).

Many focus on the impact of radiation from the SMBH at the center of the Milky Way, Sgr A*. A. Balbi & F. Tombesi (2017) focus on quantifying this atmospheric mass loss and the biological damage caused by X-ray to ultraviolet (XUV) radiation produced during Sgr A*'s peak phases of activity, which was further refined by W. Ishibashi (2024). Planetary atmospheric evaporation due to XUV radiation has also been considered on a much larger scale, with percentages of elemental loss quantified for all planets in the Universe (J. C. Forbes & A. Loeb 2018). Subsequently, this work was generalized to predict that radiation alone can result in induced atmospheric mass loss and erosion for Earth-like planets (EPs) in other galaxies (A. M. Wisłocka et al. 2019). Building on the earlier models, which did not account for atmospheric chemistry, K. I. Sippy et al. (2025) showed that the presence of sufficient atmospheric $O_2$ may help trigger the formation of an ozone layer that reduces biological damage.

While much of the attention has been on Sgr A*, limited efforts have been made to understand the effects of AGN radiation on habitable worlds in other galaxies (J. C. Forbes & A. Loeb 2018; M. Lingam 2019; A. M. Wisłocka et al. 2019). These studies have concluded that AGN activity, particularly during high-luminosity accretion phases, can induce significant atmospheric loss and biological damage on EPs, especially within the central kiloparsecs of a given galaxy. It should be observed at this juncture that radiation from AGN does not exclusively have negative consequences, because it may drive prebiotic chemistry (M. Lingam et al. 2019; C. Liu et al. 2020). However, radiation is not the only characteristic of AGN that presents a threat to habitability.

AGN winds and jets are energetic phenomena associated with the central regions of galaxies hosting active BHs. Jets are highly collimated outflows moving at relativistic speeds (M. L. Lister et al. 2021), resulting in their small angular span across a region of the host galaxy. Because of this, they have not yet been extensively considered to influence habitability. In contrast, AGN winds are broad, multidirectional outflows driven by the accretion process. One notable class, ultrafast outflows (UFOs), can reach velocities of $\sim 0.1c$ (e.g., F. Tombesi et al. 2010a, 2011, 2014, 2015; J. Gofford et al. 2013; G. Chartas et al. 2021) and are observed in both broad absorption line quasars and some Seyfert galaxies (e.g., R. J. Weymann et al. 1991; P. C. Hewett & C. B. Foltz 2003; X. Xu et al. 2019; A. L. Rankine et al. 2020).

In light of the above discussion, ionizing radiation (which includes high-energy particles) can alter surface habitability and drive long-lasting chemical changes in the interstellar medium (ISM), with effects persisting well after AGN activity ceases. However, there are only two specific publications discussing the influence of AGN winds (A. Ambrifi et al. 2022; S. Heinz 2022), which only consider the effects of a SMBH with properties reflecting those of Sgr A*. The effects on the planetary atmosphere can include atmospheric escape and ozone depletion caused by heating due to AGN winds. A. Ambrifi et al. (2022) considered these effects for Earth-like atmospheres in the Milky Way, fixing the mass of the BH examined to that of Sgr A*. This study constrains the maximum distance within the Milky Way at which the AGN wind-induced effects remain significant, finding that this limit may extend up to approximately 1 kpc. The results of this publication were then used to form a comparison between the influence of AGN winds and stellar winds on an exoplanet atmosphere (S. Heinz 2022).

Tidal disruption events (TDEs) may also contribute to galactic habitability, however their effects pertain over a much shorter timescale in contrast to the AGN phase and thus have a much smaller impact on a planetary atmosphere (E. Pacetti et al. 2020). Additionally, the emissions from TDEs are much softer than that of AGN (K. Auchettl et al. 2018).

### 1.2. This Paper

The general outline of the paper is as follows. First, in Section 2, we pinpoint areas of deficiency in the field and elucidate potential avenues for future research to address these gaps, while also outlining the specific objectives of this work. This section also describes the models and input parameters necessary to execute theoretically accurate results for AGN wind and atmospheric impacts. In the following sections, Section 3 and Section 4, we present the obtained results alongside a comprehensive discussion of their implications. Finally, in Section 5, we conclude with a concise overview of the significance of the topic and offer a summary of the findings, highlighting their relevance to the impacts of AGN and the habitability of exoplanets across galactic scales in general.

## 2. Methods

Planetary atmospheres are influenced by many external factors including various types of radiation such as XUV, as well as jets and winds from AGN. There is a lack of research regarding the specific effects of winds from AGN. Additionally, the properties of the SMBH in a galaxy may directly affect the habitability of exoplanets in a different manner than what has been studied for terrestrial worlds in the Milky Way.

Right now we have a basic understanding of the influences of AGN winds on EPs for Sgr A* in the Milky Way (A. Ambrifi et al. 2022), but it is crucial to explore the effects of winds on EPs in other galaxies that have a wide range of different masses.

Given the observed diversity in SMBH masses and AGN properties across galaxies, it is essential to examine how variations in BH mass, planetary galactocentric distance, radiative efficiency, and relative wind velocity influence the habitable zone of an exoplanet. Additionally, the effects of atmospheric depletion mechanisms outside of energy- and momentum-driven ones require intricate simulations that have not yet been executed.

Although the effects of AGN winds on the ability of planets to retain and sustain an atmosphere conducive to life seems to be greatly underresearched compared to the radiation impacts, we remark in passing that there are essentially no publications discussing the relation between jets and habitable worlds.

Since it is only possible to thoroughly consider one parameter at a time, the intention of the following material is to understand the effects of SMBH mass variability on exoplanets. As a means for comparison to those derived for the Milky Way, results will be reproduced at multiple SMBH masses for atmospheric heating, probable velocity of molecules in the atmosphere, energy- and momentum-driven atmospheric mass loss, ozone depletion, and the timescale required for 90% ozone depletion.

### 2.1. Model Description

The original code from A. Ambrifi et al. (2022), which models a single BH mass, was adapted to generate an array of BH masses.

The specific class of winds, UFOs, is modeled via the approach described in K. A. Pounds et al. (2003) under the assumption of spherical symmetry (e.g., S. Laha et al. 2021), and we further assume that the wind speed remains approximately constant.

It should be noted that UFOs interact with the ISM by generating shocks that transfer momentum and kinetic energy, giving rise to postshock winds. In the energy-driven scenario, the kinetic power of the UFO is largely conserved and transferred to the postshock wind. In contrast, in the momentum-driven case, a significant portion of the UFO's energy is radiated away soon after the shock occurs, with only the momentum translating to the postshock wind (A. Ambrifi et al. 2022).

These two cases represent limiting regimes between which a wide range of intermediate possibilities may occur. The transition between energy- and momentum-driven coupling of the AGN wind with the ISM is characterized by the critical cooling radius, $R_C \sim 500 M_8^{1/2} \sigma_{200}$ pc, where the radiative cooling time of the shocked gas becomes comparable to the flow time (A. King & K. Pounds 2015). For radii smaller than $R_C$, efficient cooling confines the coupling to a momentum-driven phase, whereas for $R > R_C$, the flow evolves adiabatically and becomes energy driven. Consequently, momentum-driven winds are expected to dominate only within the inner few hundred parsecs, while energy-driven outflows can extend to kiloparsec and galactic scales. This radius defines the region where momentum-driven interaction dominates due to efficient radiative cooling, but does not represent a strict limit on the radial extent of the outflow. Once formed, a momentum-driven wind can continue to propagate through the host galaxy, although with significantly lower coupling efficiency and large-scale impact compared to an energy-driven outflow.

An intriguing intermediate case is presented by the galaxy-scale, extended narrow-line regions (NLRs) observed in some AGN (e.g., K. N. Hainline et al. 2016; E. Congiu et al. 2017; A.-L. Sun et al. 2017), which are detected at distances up to $\lesssim$10 kpc from the nucleus. Such extended NLRs may initially be energy driven, later transitioning to a momentum-driven phase as radiative cooling becomes significant. This balance may also be affected by the $M_{BH}-\sigma$ relation (K. Gebhardt et al. 2000) which links the mass of the BH to the galactic velocity dispersion. At present, there is no clear consensus as to whether UFOs are predominantly energy or momentum driven (see, e.g., K. Pounds et al. 2016; M. Mizumoto et al. 2019; Y. Xu et al. 2024). In this study, both regimes are evaluated across a uniform range of galactocentric distances (0.1–100 kpc) for comparative purposes, but this physical distinction in spatial applicability is recognized.

When considering the data for the central SMBH, it is necessary to determine the characteristic luminosity. The Eddington luminosity associated with the BH is

$$L_{\rm Edd} = \frac{4\pi G c m_{\rm p}}{\sigma_{\rm T}} M_{\rm BH} \approx 3.3 \times 10^4 \left(\frac{M_{\rm BH}}{M_\odot}\right) L_\odot, \qquad (1)$$

where $G$ is the gravitational constant, $m_p$ is the mass of the proton, $\sigma_T$ is the Thomson scattering cross section, and the mass of the BH is signified by $M_{BH}$.

The Salpeter timescale can be determined after calculating the Eddington luminosity by

$$\Delta t_{\rm Salp} = \frac{M_{\rm BH} \eta c^2}{(1-\eta) L_{\rm Edd}}, \qquad (2)$$

where $\eta$ is the radiative efficiency.

A typical AGN phase is estimated to last on the order of $10^7$–$10^9$ yr (P. Martini & D. H. Weinberg 2001; A. Marconi et al. 2004). These durations are comparable to the Salpeter timescale $t_{Salp} \sim 10^7$ yr, which characterizes the $e$-folding time for BH mass growth under Eddington-limited accretion (J. S. B. Wyithe & A. Loeb 2003).

Although individual AGN episodes may occur on shorter timescales ($\sim 10^4$–$10^5$ yr; e.g., K. Zubovas 2018; K. Zubovas et al. 2022), this study focuses on the cumulative effects of such episodes considering the Salpeter timescale as an estimate of the total AGN phase.

According to B. Czerny et al. (1997), the Eddington ratio $L/L_{Edd}$ in quasars typically falls within the range of $\sim$0.01–0.1, while Seyfert galaxies exhibit a broader distribution, spanning from $\sim$0.001 up to 0.3 or higher.

### 2.2. Input Parameters

We consider SMBHs with masses ranging from $10^6$ up to $10^{10}\, M_\odot$, consistent with the observed mass range. We adopt a typical value of the radiative efficiency of $\eta = 0.1$ (Y. Shen 2013). The characteristic velocities of UFOs, typically $0.1c$ (A. Luminari et al. 2021), and of warm absorbers ($10^4$–$10^5$ km s$^{-1}$; Z. Igo et al. 2020), are determined in F. Tombesi et al. (2013). Planetary and atmospheric parameters are held constant, modeled after Earth-like characteristics, with a planetary radius $R_p = R_\oplus$ and a density of $\rho_p \approx \rho_\oplus \approx 5.5$ g cm$^{-3}$.

Since we are considering two types of atmospheric composition, dominated by either molecular nitrogen $N_2$ or hydrogen $H_2$, the constant values of the specific heat capacities are determined using data from the NIST Standard Reference Database.[7] Following the method outlined in A. Ambrifi et al. (2022), the corresponding specific heat capacities were determined to be $C_{N_2} \approx 1.3 \times 10^7$ erg g$^{-1}$ K$^{-1}$ for molecular nitrogen and $C_{H_2} \approx 1.83 \times 10^8$ erg g$^{-1}$ K$^{-1}$ for molecular hydrogen. As demonstrated later in Section 3.2, the effect of the molar mass is linear in our model.

## 3. Atmospheric Heating and Escape Due to Active Galactic Nucleus Winds

### 3.1. Sample Selection of Active Galaxies

As a means for comparison, plots of galaxies with central galactic SMBHs at different orders of magnitude in mass have been modeled for each dataset. The sample galaxies chosen are NGC 1068 ($M_{\rm BH} = 1.3 \times 10^7 M_\odot$; J.-M. Wang et al. 2020), NGC 5128 ($M_{\rm BH} = 5.5 \times 10^7 M_\odot$; M. Cappellari et al. 2009), 3C 390.3 ($M_{\rm BH} = 2.8 \times 10^8 M_\odot$; S. G. Sergeev et al. 2016), M104 ($M_{\rm BH} = 1 \times 10^9 M_\odot$; J. Kormendy et al. 1996), M87 ($M_{\rm BH} = 6.5 \times 10^9 M_\odot$; K. Akiyama et al. 2019), and finally OJ 287 primary ($M_{\rm BH} = 1.8 \times 10^{10} M_\odot$; M. J. Valtonen et al. 2016). Additionally, the plots for Sgr A$^*$ have been reproduced identically to those in A. Ambrifi et al. (2022) to allow for direct comparison, noting the adoption of a slightly updated SMBH mass of $4.3 \times 10^6 M_\odot$ (Event Horizon Telescope Collaboration et al. 2022) consistent with more recent dynamical measurements.

[7] NIST Chemistry WebBook: https://webbook.nist.gov/chemistry/.

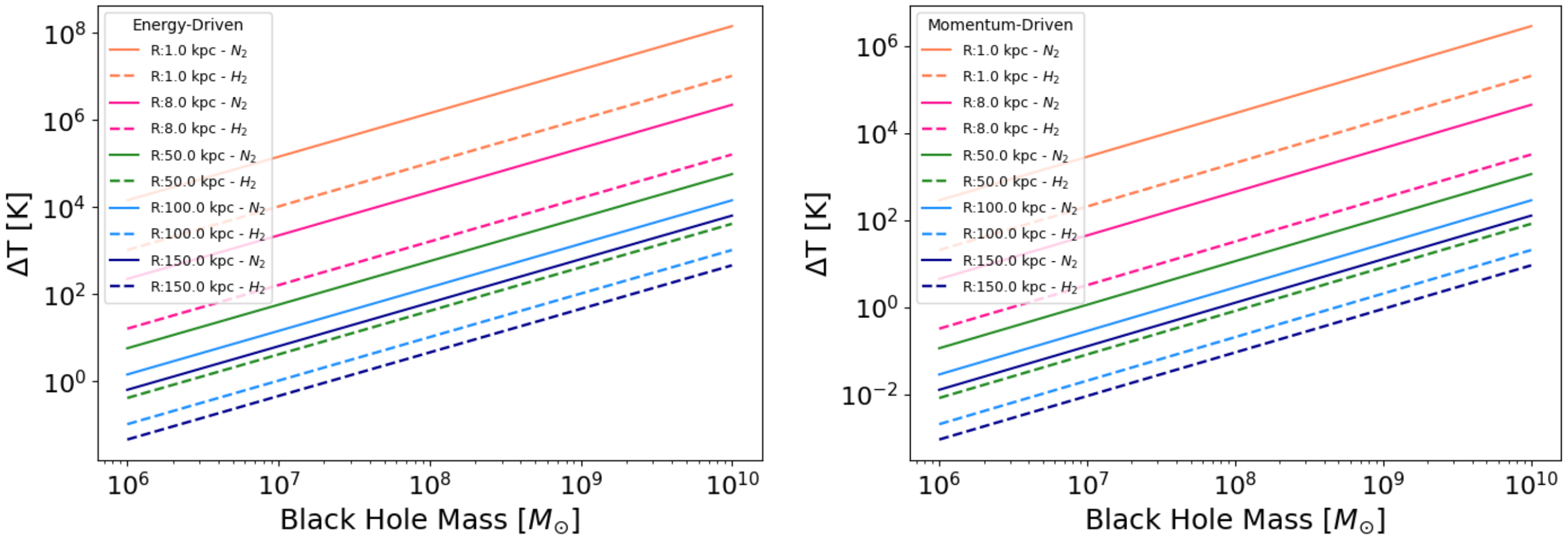

**Figure 1.** Increase in atmospheric temperature caused by AGN winds as a function of the mass of the central galactic SMBH in solar masses. The left panel shows the energy-driven case, while the right one shows the effect of momentum-driven winds. The lines represent the distance $R$ from the galactic center (in kiloparsecs). The labels $N_2$ and $H_2$ indicate the main element of planetary atmospheric composition, molecular nitrogen or hydrogen, respectively.

The small sample of active galaxies was chosen to span several orders of magnitude in SMBH mass, without selecting other parameters. Some of these sources, notably NGC 1068 (M. Mizumoto et al. 2019) and 3C 390.3 (F. Tombesi et al. 2010b), have been reported to host UFOs, while others are classified as radio-loud AGN. The primary objective was to explore the dependence on SMBH mass, and the potential effects of radio-quiet versus radio-loud AGN will be considered in future work. It is also worth noting that recent studies have demonstrated the ubiquitous presence of winds, in particular UFOs, in both radio-quiet and radio-loud AGN, making the discussion of UFO effects more broadly applicable (e.g., S. Mestici et al. 2024). In all cases, we focus specifically on UFOs and consider only the planets located within the region of the wind.

### 3.2. *Atmospheric Heating and Escape*

The interaction between AGN winds and a planetary atmosphere can result in significant energy deposition, leading to a rise in atmospheric temperature. The first plot (Figure 1) illustrates the atmospheric temperature change ($\Delta T$) as a function of BH mass, for different values of galactocentric distance $R$. Figure 2 depicts the atmospheric temperature increase, shown as a function of distance from the galactic center rather than SMBH mass. Each panel represents a different example galaxy, with SMBH masses that vary from one case to the next. These results are derived under the simplified assumption that all incident AGN wind energy is deposited into atmospheric heating.

As described in A. Ambrifi et al. (2022), the kinetic energy associated with the outflow can be written as

$$\dot{\epsilon}_{\rm k} \equiv \frac{1}{2}\dot{M}_{\rm out} v_{\rm out}^2 \approx 0.05 \lambda_{\rm Edd} L_{\rm Edd}\left(\frac{v_{\rm out}}{0.1c}\right), \tag{3}$$

where $\dot{M}_{\rm out}$ is the mass outflow rate and $v_{\rm out}$ is the outflow velocity. For a UFO, $v_{\rm out} \sim 0.1c$. In the second relation, the Eddington ratio parameter is taken to be $\lambda_{\rm Edd} \approx 1$, consistent with observational estimates (A. Marconi et al. 2004; C. L. Steinhardt & M. Elvis 2010; M. Bischetti et al. 2017; J. Aird et al. 2018).

In general, $\epsilon$ represents the coupling efficiency, or the fraction of AGN luminosity transformed into kinetic energy of the outflowing wind, such that

$$\dot{\epsilon}_{\rm k} = \epsilon L_{\rm Edd}. \tag{4}$$

As noted previously in Section 2.1, for the energy-driven case, we assume the kinetic power of the postshock wind is equal to that of the UFO. Therefore the kinetic power of the energy-driven postshock wind is simplified to

$$\dot{\epsilon}_{\rm k,ed} \approx \dot{\epsilon}_{\rm k} \approx 0.05 L_{\rm Edd}. \tag{5}$$

The kinetic power of the momentum-driven wind is determined by the velocity of the postshock wind $v_{\rm ps}$, rather than the initial UFO velocity, $v_{\rm out}$. As shown in detail by A. Ambrifi et al. (2022), it can be expressed as

$$\dot{\epsilon}_{\rm k,md} \approx \frac{1}{2}\frac{\lambda_{\rm Edd} L_{\rm Edd}}{c} v_{\rm ps}, \tag{6}$$

where $v_{\rm ps} \approx 1000\ {\rm km\ s^{-1}}$ (J. Debuhr et al. 2012; C.-A. Faucher-Giguère & E. Quataert 2012; K. Zubovas & A. King 2012; A. King & K. Pounds 2015; F. Tombesi et al. 2015).

This can be further simplified using the conservation of momentum:

$$\dot{\epsilon}_{\rm k,md} \approx 0.001 \lambda_{\rm Edd} L_{\rm Edd}\left(\frac{v_{\rm ps}}{1000\ {\rm km\ s^{-1}}}\right) \approx 0.001 L_{\rm Edd}, \tag{7}$$

implying $\epsilon \approx 0.05$ for energy-driven winds and $\approx 0.001$ for momentum-driven winds.

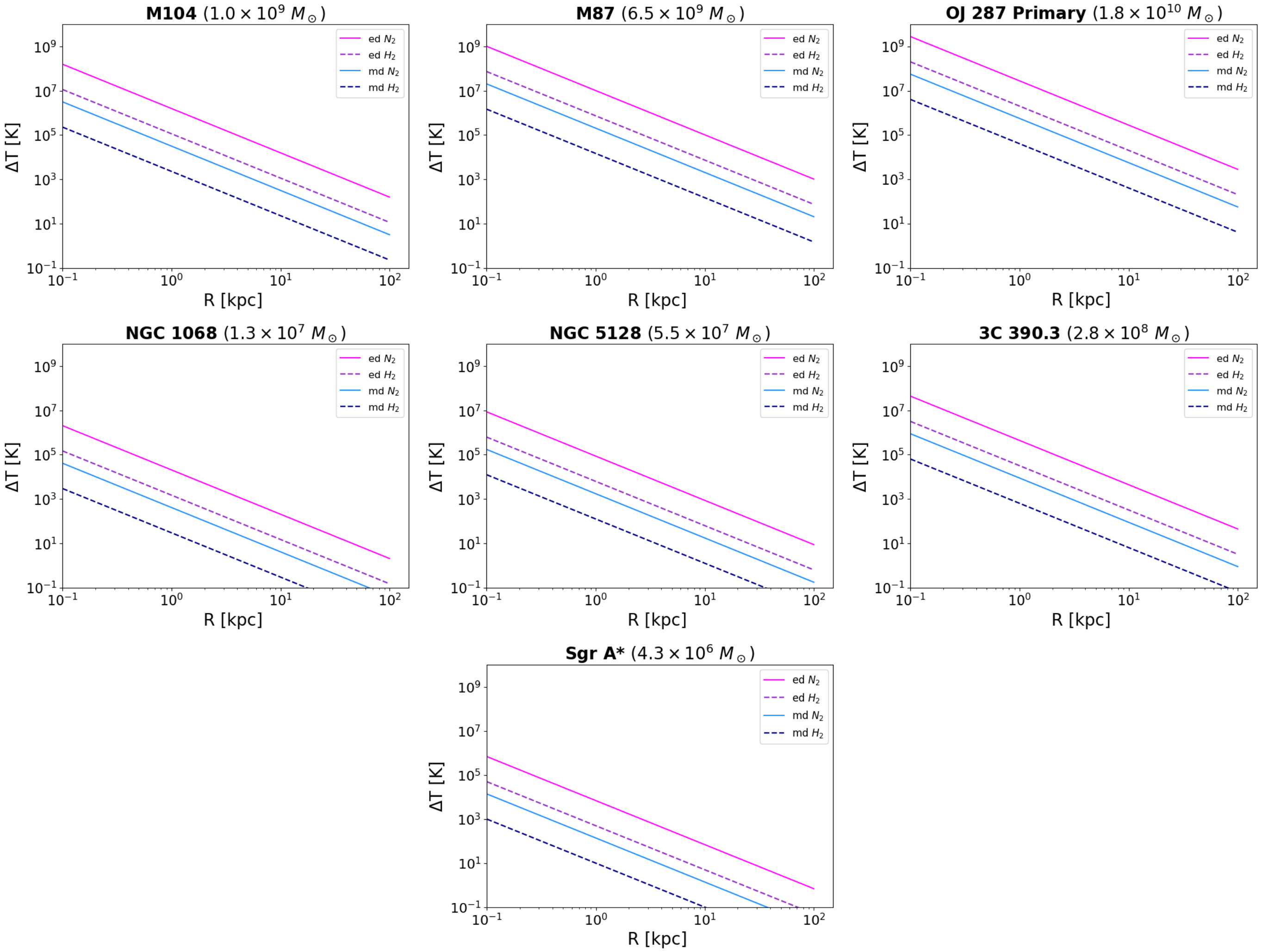

**Figure 2.** Increase in atmospheric temperature caused by energy- and momentum-driven AGN winds as a function of the distance to the galactic center (in kiloparsecs). The labels $N_2$ and $H_2$ indicate the main element of planetary atmospheric composition, molecular nitrogen or hydrogen, respectively.

An upper bound to the atmospheric temperature increase can be estimated by evaluating the total incident energy from the outflow, obtained by multiplying the total kinetic power $\dot{\epsilon}_k$ distributed over a spherical surface by the Salpeter time, divided by the atmospheric mass and specific heat capacity. The full derivation is presented in A. Ambrifi et al. (2022), and the general expression for the change in atmospheric temperature is given by

$$\Delta T = \frac{\epsilon L_{\mathrm{Edd}} \Delta t_{\mathrm{Salp}}}{4 M_{\mathrm{atm},\oplus} C} \left(\frac{R_{\mathrm{p}}}{R}\right)^2, \tag{8}$$

where $M_{\mathrm{atm},\oplus}$ is the atmospheric mass (assumed to be Earth-like, $M_{\mathrm{atm},\oplus} \approx 5.1 \times 10^{21}$ g), $C$ is the specific heat capacity, and $R_{\mathrm{p}}/R$ is the ratio of planetary radius to AGN distance. In this simplified model, the total incident energy is assumed to heat the atmosphere uniformly.

As shown in figures, larger SMBHs result in greater atmospheric heating, while increasing galactocentric distance significantly mitigates this effect, both results aligning with intuitive physical expectations. This trend follows directly from Equation (8), which shows that the temperature scales as $\Delta T \propto M_{\mathrm{BH}}/R^2$. However, in Figures 1 and 2, there is a temperature regime between $10^6$ and $10^8$ K where our results are nonphysical. As shown in Figure 2, the energy-driven scenario for M104 reaches atmospheric temperatures of roughly $10^6$ K at 1 kpc. In even more massive systems, such as OJ 287 primary, the heating becomes extreme for distances up to 5 kpc.

To evaluate the implications of this heating, the most probable molecular velocity ($v_{\mathrm{mp}}$) of atmospheric particles (E. H. E. H. Kennard 1938) is calculated via

$$v_{\mathrm{mp}} = \sqrt{\frac{2 k_{\mathrm{B}} T'}{m_{\mathrm{s}}}}, \tag{9}$$

where the new temperature is $T' = T_0 + \Delta T$ with $T_0 \approx 273$ K, $m_{\mathrm{s}}$ is the mass of a single molecule (e.g., $N_2$ or $H_2$), and $k_{\mathrm{B}}$ is the Boltzmann constant. These velocities are shown in Figure 3, corresponding to the energy-driven and momentum-driven wind models.

For atmospheric escape, the model assumes that the energy deposited in the atmosphere is transmuted into the kinetic energy of atmospheric particles. If a substantial fraction of molecules attain velocities greater than the planet's escape velocity, significant atmospheric loss via thermal escape may

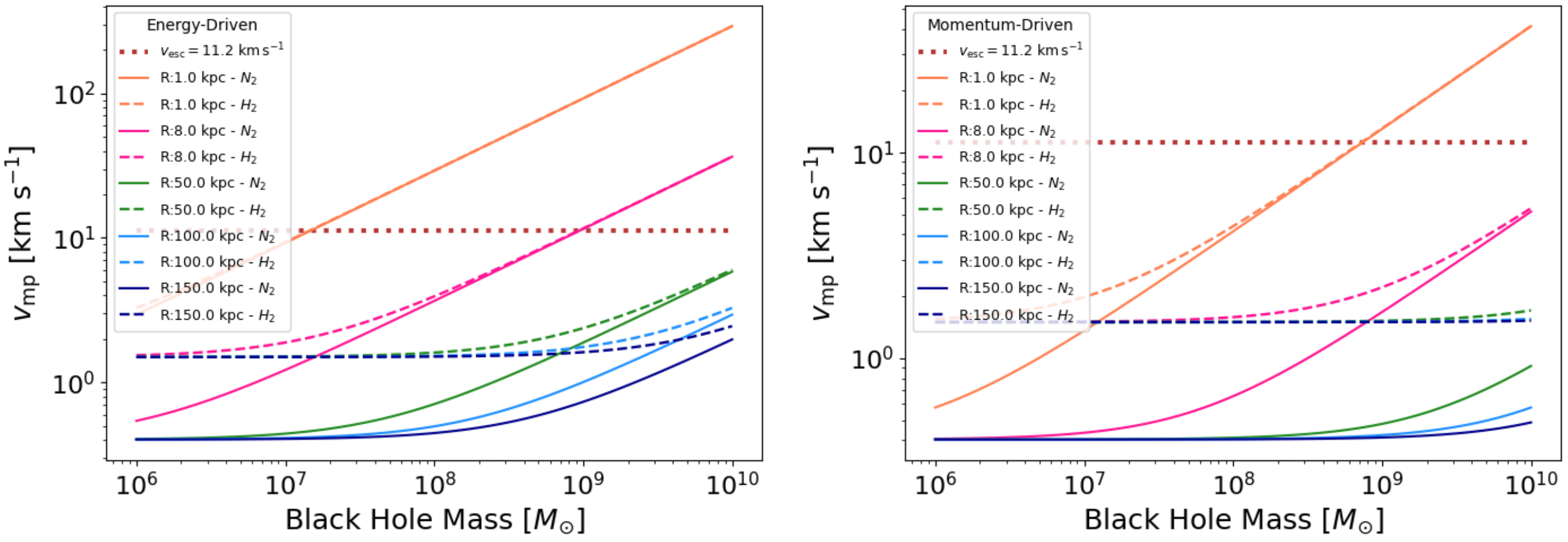

**Figure 3.** Most probable velocity of molecules in the planetary atmosphere ($v_{mp}$) by energy- and momentum-driven AGN winds as a function of the mass of the central galactic SMBH in solar masses. The left panel shows the energy-driven case, while the right one shows the effect of momentum-driven winds. The lines represent the distance $R$ from the galactic center (in kiloparsecs). The labels $N_2$ and $H_2$ indicate the main element of planetary atmospheric composition, molecular nitrogen or hydrogen, respectively. The horizontal line represents the escape velocity of the Earth ($v_{esc} \approx 11.2$ km s$^{-1}$).

occur. The escape velocity for Earth is given by

$$v_{esc} = \sqrt{\frac{2GM_\oplus}{R_\oplus}} \approx 11.2 \text{ km s}^{-1}, \tag{10}$$

and is represented as a solid horizontal line in the velocity plots. These figures allow one to identify, for each SMBH mass and galactocentric distance, whether molecular nitrogen or hydrogen can escape the EP's gravity. For example, in the energy-driven case (Figure 3) at $R = 1$ kpc, both $N_2$ and $H_2$ can exceed the escape velocity for SMBH masses as low as $10^7 M_\odot$. In the momentum-driven scenario (Figure 3), escape is only feasible at close distances ($R = 1$ kpc) for SMBH masses approaching $10^9 M_\odot$. At distances greater than this, the effect of momentum-driven winds on the most probably velocity of atmospheric particles is negligible, as the velocity would only exceed that of the Earth's at BH masses much greater than the considered range.

In general, $v_{mp}$ increases with SMBH mass and decreases with galactocentric distance, as expected from Equation (9), which shows that $v_{mp} \propto M_{BH}^{1/2}/R$. Each plot includes a range of distances from 1 kpc to 150 kpc and considers two molecular species: nitrogen ($N_2$), representative of modern Earth's atmosphere, and hydrogen ($H_2$), characteristic of super-Earth atmospheres. These hydrogen-rich atmospheres are expected to be more common and potentially habitable, though they typically possess lower atmospheric mass (L. T. Elkins-Tanton & S. Seager 2008; S. Seager et al. 2013, 2020; N. Madhusudhan et al. 2021; A. Ambrifi et al. 2022; M. Lingam & A. Balbi 2024).

The difference between these two species becomes more pronounced at larger radial distances from the BH, as seen in Figures 3 and 4. This arises because, as $R$ increases, the UFO-induced atmospheric heating diminishes ($\Delta T \to 0$), and the most probable velocity depends more strongly on molecular mass. As a result, lighter species, including $H_2$, maintain higher velocities than heavier molecules such as $N_2$.

We extend our analysis to galactocentric distances as large as 150 kpc to encompass the full range of plausible environments impacted by the central SMBH. Observational evidence supports the relevance of such scales: the stellar halo of M87 extends to ~150 kpc (M. Doherty et al. 2009; A. Longobardi et al. 2015), and NGC 5128 reaches similar projected distances (D. Crnojević et al. 2016). Additionally, massive satellite galaxies can reside at comparable distances from their host, making it important to assess atmospheric vulnerability even in these outer galactic regions.

These trends are quantified in Table 1, where the escape conditions are determined based on when $v_{mp}$ exceeds $v_{esc}$ in the plotted results (Figure 4). Both the figure and Table 1 confirm that atmospheric escape via AGN winds becomes increasingly unlikely with galactocentric distance, but grows more severe with SMBH mass.

These results indicate that AGN wind-driven atmospheric escape is most pronounced in the inner galactic regions of galaxies hosting high-mass SMBHs. Energy-driven winds, in particular, have a far greater capacity to drive atmospheric loss than momentum-driven ones, posing a significant hazard to planetary atmospheres, especially those of terrestrial or super-Earth planets orbiting within a few kiloparsecs of an AGN.

We note that there is a regime where Figures 1 and 2 give results that are not strictly physical. This is in the regime of temperatures of tens of thousands of kelvin or above, where the most probable velocity $v_{mp}$ exceeds the escape velocity $v_{esc}$. In this regime, catastrophic atmospheric loss will occur and it is reasonable to conclude that the atmosphere would be completely stripped and the assumptions of our simplified heating model would no longer hold. These threshold values are obtained by taking Equation (9), substituting $T' = 273$ K$+ \Delta T$, and solving for $\Delta T$ under the condition $v_{mp} = v_{esc}$ for each molecular species. For example, using $v_{esc} = 11.2$ km s$^{-1}$, $v_{mp}$ for molecular hydrogen surpasses Earth's escape velocity at a temperature of ~41,000 K ($\Delta T \approx$ 41,000 K), while molecular nitrogen crosses this threshold at ~567,000 K ($\Delta T \approx$ 567,000 K). Even at somewhat lower temperatures, a significant fraction of particles in the high-velocity tail of the Maxwellian distribution can escape, so that over many thermal timescales (~a few years for an Earth-like atmosphere; e.g., S. E. Schwartz 2007), atmospheric escape would still be severe. In addition to just atmospheric escape, such extreme levels of heating would naturally have other profound impacts such as ionization of the atmosphere.

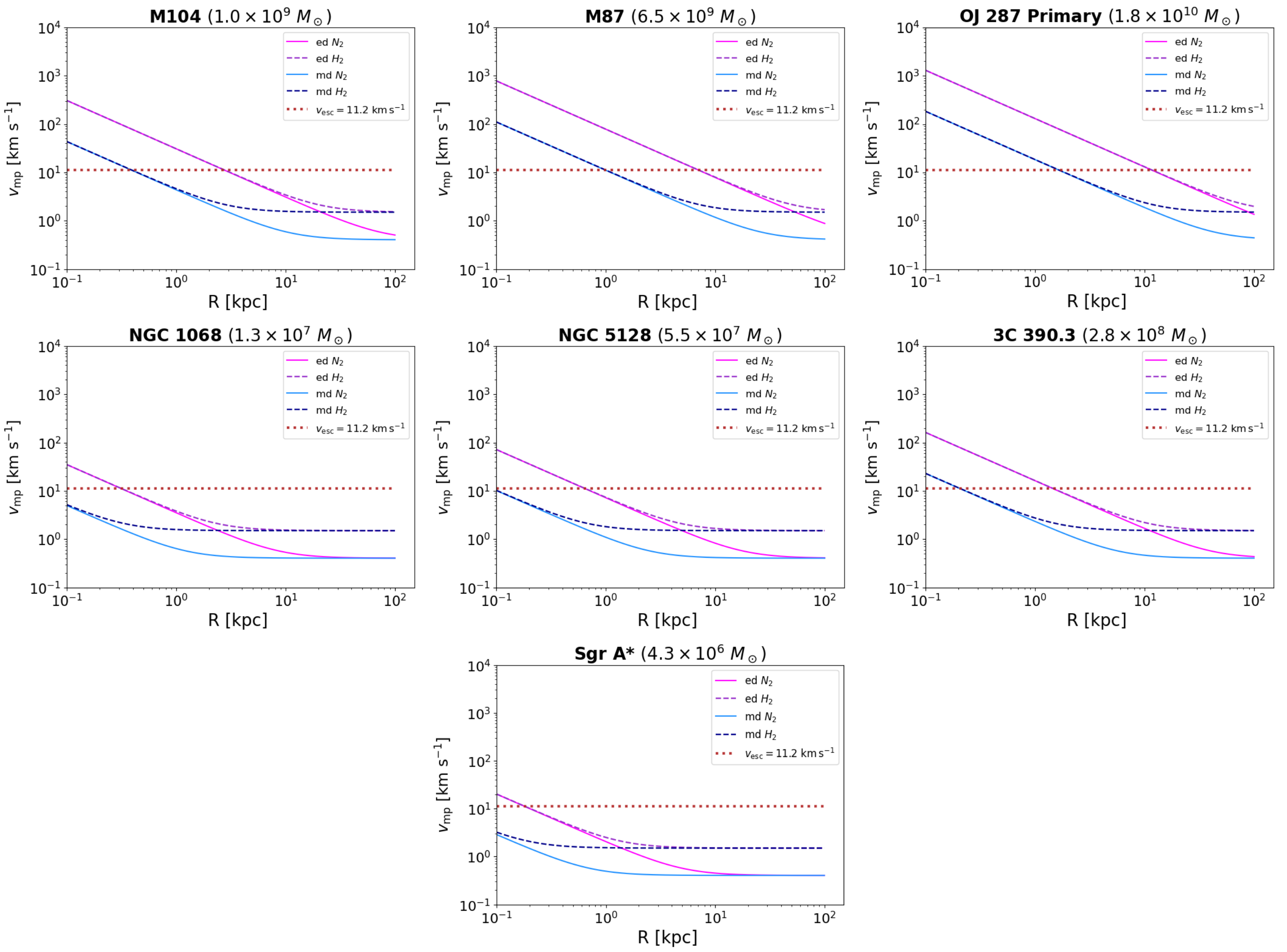

**Figure 4.** Most probable velocity of molecules in the planetary atmosphere ($v_{\rm mp}$) by momentum- and energy-driven AGN winds as a function of the distance to the central galactic SMBH (in kiloparsecs). The labels $N_2$ and $H_2$ indicate the main element of planetary atmospheric composition, molecular nitrogen or hydrogen, respectively. The horizontal line represents the escape velocity of the Earth ($v_{\rm esc} \approx 11.2$ km s$^{-1}$).

### 3.3. Atmospheric Mass Loss

Next, we adopt an alternative mechanism for atmospheric escape, based on the well-established framework of energy-limited hydrodynamic escape (D. C. Catling & J. F. Kasting 2017; J. E. Owen 2019; G. Gronoff et al. 2020; M. Lingam & A. Loeb 2021; D. Kubyshkina et al. 2026). In this framework, the incident energy is converted into kinetic energy of atmospheric particles, allowing them to escape. Hydrodynamic escape driven by AGN activity has previously been investigated by A. Balbi & F. Tombesi (2017), J. C. Forbes & A. Loeb (2018), and A. M. Wisłocka et al. (2019). Following this approach, we adopt the energy-limited formulation utilized by A. Ambrifi et al. (2022), in which the AGN wind energy replaces the electromagnetic energy as the driver of atmospheric escape. The reason we transition to this framework is because it has been shown in Section 3.2 that the particles' mean energy can exceed the gravitational escape energy. When this regime is entered, it is well understood that atmospheric escape will not just be restricted to the high-energy particles, but to the bulk; in other words, the atmosphere functions as a fluid and experiences bulk escape through the mechanism of hydrodynamic atmospheric escape (D. Kubyshkina et al. 2026, Section 2.1.2). Hence, as the existence of this regime has been shown to be viable, it is necessary to model it.

Assuming all of the wind energy is transformed to kinetic energy of escaping particles, the atmospheric mass loss can be found as

$$M_{\rm lost} = \frac{3}{16\pi G\rho_{\rm p}}\frac{\epsilon L_{\rm Edd}\Delta t_{\rm Salp}}{R^2}, \tag{11}$$

where $R$ is the radial distance from the galactic center.

Figure 5 represents the atmospheric mass loss due to energy-driven and momentum-driven wind-mediated escape. Both results depict a linear relationship between the mass-loss fraction and the BH mass. In each case, as the radius increases, the mass-loss fraction declines. As the mass of the BH grows, all mass-loss fractions increase proportionally. The only difference between the energy-driven and momentum-driven scenarios is that the overall mass fractions have smaller values in the momentum-driven case.

For BH masses greater than $10^8\,M_\odot$, the ratio of atmospheric mass lost to Earth's present atmospheric mass ($M_{\rm lost}/M_{\rm atm,\oplus}$) increases significantly. As shown in Figure 6,

**Table 1**
Potential Maximum Galactic Distances Where the Effects of Ultrafast Outflows from Active Galactic Nuclei Remain Significant

| Effect | Mass ($M_\odot$) | Momentum-driven Case (kpc) | Energy-driven Case (kpc) |
|---|---|---|---|
| Atmospheric escape arising from thermal heating | $4.3 \times 10^6$ | N/A | ∼0.2 |
| … | $1.3 \times 10^7$ | N/A | ∼0.3 |
| … | $5.5 \times 10^7$ | <0.1 | ∼0.7 |
| … | $2.8 \times 10^8$ | ∼0.2 | ∼1.5 |
| … | $1.0 \times 10^9$ | ∼0.4 | ∼3 |
| … | $6.5 \times 10^9$ | ∼1 | ∼7 |
| … | $1.8 \times 10^{10}$ | ∼1.5 | ∼12 |
| Energy-limited hydrodynamic-like atmospheric escape | $4.3 \times 10^6$ | <0.1 | ∼0.4 |
| … | $1.3 \times 10^7$ | ∼0.1 | ∼0.7 |
| … | $5.5 \times 10^7$ | ∼0.3 | ∼1.5 |
| … | $2.8 \times 10^8$ | ∼0.6 | ∼3 |
| … | $1.0 \times 10^9$ | ∼1 | ∼6 |
| … | $6.5 \times 10^9$ | ∼3 | ∼15 |
| … | $1.8 \times 10^{10}$ | ∼5 | ∼25 |
| Major ozone depletion (90% loss) due to nitrogen oxide formation | $4.3 \times 10^6$ | ∼0.11 | ∼1 |
| … | $1.3 \times 10^7$ | ∼0.15 | ∼1.7 |
| … | $5.5 \times 10^7$ | ∼0.3 | ∼3.7 |
| … | $2.8 \times 10^8$ | ∼1.1 | ∼8 |
| … | $1.0 \times 10^9$ | ∼2 | ∼16 |
| … | $6.5 \times 10^9$ | ∼6 | ∼40 |
| … | $1.8 \times 10^{10}$ | ∼9 | ∼67 |

**Note.** The term "N/A" is used when the effects in question are substantial only up to relatively negligible distances. For comparison, the modeled SMBH masses correspond to observed galaxies spanning several orders of magnitude: Sgr A* ($M_{BH} = 4.3 \times 10^6 M_\odot$; Event Horizon Telescope Collaboration et al. 2022), NGC 1068 ($M_{BH} = 1.3 \times 10^7 M_\odot$; J.-M. Wang et al. 2020), NGC 5128 ($M_{BH} = 5.5 \times 10^7 M_\odot$; M. Cappellari et al. 2009), 3C 390.3 ($M_{BH} = 2.8 \times 10^8 M_\odot$; S. G. Sergeev et al. 2016), M104 ($M_{BH} = 1 \times 10^9 M_\odot$; J. Kormendy et al. 1996), M87 ($M_{BH} = 6.5 \times 10^9 M_\odot$; K. Akiyama et al. 2019), and OJ 287 primary ($M_{BH} = 1.8 \times 10^{10} M_\odot$; M. J. Valtonen et al. 2016). The critical distances reported in each column correspond to the galactocentric radii at which the respective atmospheric effects become significant. These distances are determined by distinct physical criteria: atmospheric escape from thermal heating is identified where the most probable velocity ($v_{mp}$) equals the planetary escape velocity ($v_{esc}$); energy-limited hydrodynamic-like atmospheric escape occurs where the cumulative atmospheric mass loss ($M_{lost}$) equals Earth's present atmospheric mass ($M_{atm,\oplus}$); and major ozone depletion (90% loss) due to nitrogen oxide formation is defined by the intersection of the depletion timescale with NO residence time ($\Delta t \approx 4$ yr).

when this ratio reaches values of three or higher, substantial depletion occurs even at moderate galactocentric distances, e.g., ∼2 kpc in 3C 390.3 and ∼14 kpc in OJ 287 primary.

Figure 5 further illustrates that $M_{lost}/M_{atm,\oplus}$ well exceeds $10^2$ for BH masses above $10^9 M_\odot$ at a distance of 1 kpc in the energy-driven case. This implies that even atmospheres as massive as Venus's (∼90–100 times that of Earth) would be fully stripped under such conditions. While more massive planets, such as super-Earths, may have deeper gravitational wells capable of retaining their atmospheres more effectively, their long-term viability is still uncertain. A variety of publications have shown that X-ray-driven evaporation can erode substantial portions of hydrogen envelopes from low-mass planets, particularly in the early active phases of stellar evolution (J. E. Owen 2019; W. Zhu & S. Dong 2021). They also suggest that even a few percent of atmospheric mass can be lost through thermal evaporation over megayear timescales, potentially leaving behind a bare rocky core. However, the driving evaporative mechanisms considered in such publications are radiative (X-ray and UV), whereas AGN winds represent a more particle- and fluid-dynamics-dominated process, more akin to solar wind interactions than stellar irradiation, whose importance is being increasingly appreciated for exoplanets (M. Lingam & A. Loeb 2019a, 2021; V. S. Airapetian et al. 2020).

The results in Table 1 for the energy-limited hydrodynamic-like atmospheric escape are determined from Figure 6. By setting the ratio of atmospheric mass loss to Earth's atmospheric mass equal to one ($M_{lost}/M_{atm,\oplus} = 1$), it becomes straightforward to determine the distance at which a planet's atmosphere would lose an amount of mass equivalent to Earth's. This distance is one of many crucial parameters for determining the habitable zone, as a planet without an atmosphere would be incapable of supporting surface life. However, even partial atmospheric loss may compromise planetary habitability. Substantial atmospheric mass loss represents a conservative lower bound for determining when an atmosphere can no longer support life, among other reasons due to the finite atmospheric pressure required to support liquid water. For instance, the ozone layer is confined to the stratosphere, roughly 15–35 km above the surface.[8] As a result, even 1% atmospheric mass loss could correspond to a significant depletion of ozone. It is, however, possible that life may persist in deep subsurface refugia (W. B. Whitman et al. 1998; C. Magnabosco et al. 2018), although such environments are unlikely to produce detectable biosignatures (S. McMahon et al. 2013; M. Lingam & A. Loeb 2019b, 2020).

It is important to note that energy-limited escape represents only one of several mechanisms responsible for atmospheric loss. In particular, we do not consider nonthermal ion escape processes (e.g., polar winds), which are known to be major contributors to atmospheric depletion on weakly magnetized planets (D. A. Brain et al. 2016; C. Dong et al. 2018a, 2018b; M. Lingam & A. Loeb 2019a, 2021; V. S. Airapetian et al. 2020). A comprehensive treatment of this issue would require advanced multispecies magnetohydrodynamic simulations, which are well beyond the scope of this study.

Lastly, we comment briefly on the timescales for atmospheric escape mediated by UFO outflows relative to that driven by stellar winds. An estimate of the latter is challenging to undertake, given that it involves a variety of factors ranging from the XUV flux, to stellar wind parameters (e.g., density, temperature, and velocity), to planetary characteristics. However, to offer one point of comparison, the current atmospheric escape rates for Mars and Earth due to stellar wind processes are roughly $\sim 10^{25}$–$10^{26}$ particles $s^{-1}$ (R. J. Lillis et al. 2017; C. Dong et al. 2018b; G. Gronoff et al. 2020; W. K. Peterson et al. 2024), which translate to an atmospheric mass equal to that of the Earth being depleted on timescales of order 10–100 Gyr. In contrast, the timescales considered for AGN-mediated escape herein correspond to that of the Salpeter time, which is of order 10–100 Myr. On the other hand, for certain planets around M dwarfs, the timescale for depleting an Earth-like atmosphere could approach $\sim$100 Myr in extreme cases with intense stellar winds (C. Dong et al. 2017; M. Lingam & A. Loeb 2018; V. S. Airapetian et al. 2020).

[8] NOAA Chemical Sciences Laboratory: https://csl.noaa.gov/assessments/ozone/2014/.

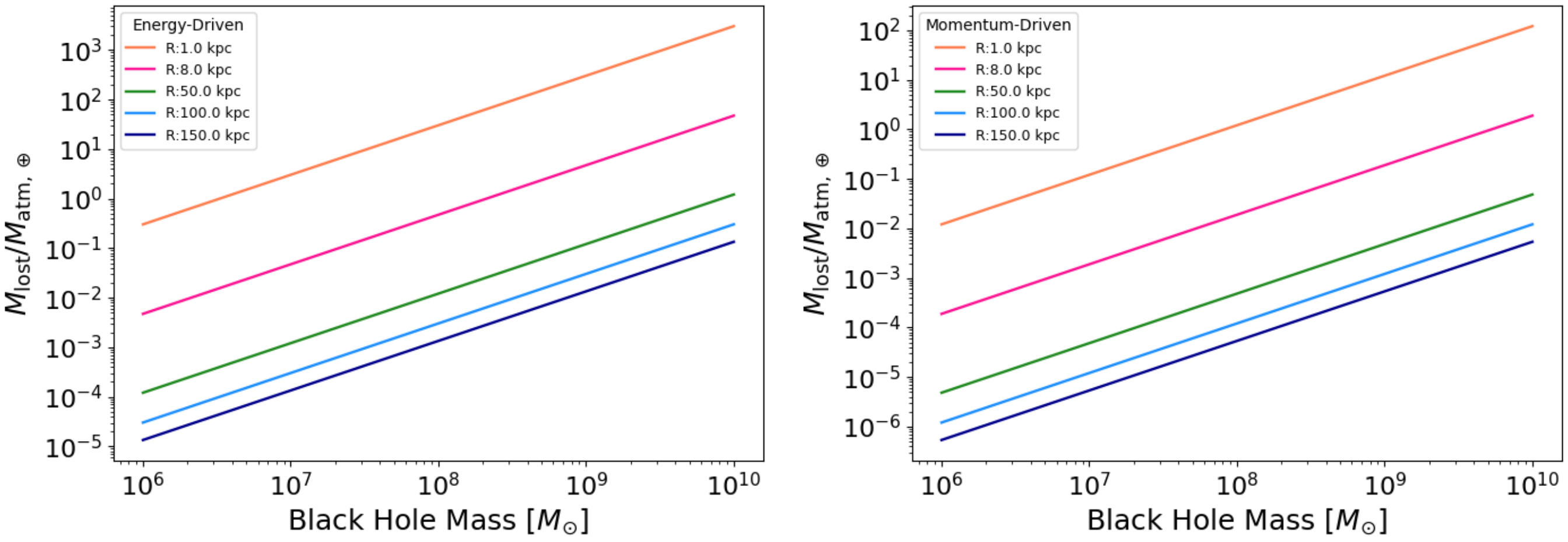

**Figure 5.** Atmospheric mass loss (relative to Earth's atmospheric mass) due to wind-mediated escape as a function of the mass of the central galactic SMBH in solar masses. The left panel shows the energy-driven case, while the right one shows the effect of momentum-driven winds. The lines represent the distance $R$ from the galactic center (in kiloparsecs).

## 4. Ozone Depletion and Consequences

### 4.1. The Impact of Energetic Particles

Energetic particles, such as those generated during solar flares, are known to initiate the formation of nitrogen oxides ($NO_x$), which catalytically destroy $O_3$ in the atmosphere (P. J. Crutzen 1971; P. J. Crutzen et al. 1975). The abundance of $NO_x$ species (NO and $NO_2$) can increase significantly in response to an enhanced flux of high-energy particles (E. S. Cramer et al. 2017; A. Ambrifi et al. 2022). These particles ionize the atmosphere, generating ion pairs that interact with molecular nitrogen ($N_2$), leading to its dissociation (P. J. Crutzen et al. 1975; L. R. Dartnell 2011). The resulting free nitrogen atoms initiate reactions that both create and recycle nitric oxide (NO):

$$\mathrm{N} + \mathrm{O_2} \rightarrow \mathrm{NO} + \mathrm{O}, \tag{12}$$

$$\mathrm{N} + \mathrm{NO} \rightarrow \mathrm{N_2} + \mathrm{O}. \tag{13}$$

The NO produced via these pathways then catalyzes ozone depletion through well-known reaction cycles:

$$\mathrm{NO} + \mathrm{O_3} \rightarrow \mathrm{NO_2} + \mathrm{O_2}, \tag{14}$$

$$\mathrm{NO_2} + \mathrm{O} \rightarrow \mathrm{NO} + \mathrm{O_2}. \tag{15}$$

In parallel, solar proton events can also lead to the production of nitrous oxide ($N_2O$), a greenhouse gas with a warming potential roughly 300 times that of carbon dioxide ($CO_2$) on a 100 yr timescale (C. Voigt et al. 2017). This warming potential suggests that $N_2O$ may have contributed to greenhouse conditions on the early Earth and could also affect the climate and habitability of exoplanets exposed to similar particle fluxes (D. E. Canfield et al. 2010; M. Lingam 2019).

### 4.2. Depletion of Ozone

The ozone depletion modeling methodology is comprehensively detailed in A. Ambrifi et al. (2022), drawing on J. Ellis & D. N. Schramm (1995). Here, we expand on this approach to provide additional details essential for extending the model to a broader range of masses.

First, we consider the nitrogen monoxide production rate due to an increased flux of cosmic particles:

$$R_{\mathrm{NO}} = R_0 \frac{\Phi}{\Phi_0} \frac{10 + y_0}{10 + y} \text{ molecules cm}^{-2}\text{ yr}^{-1}, \tag{16}$$

where $R_0$ is the rate at which NO is generated by the ambient flux of cosmic rays (CRs), $\Phi$ is the energy flux carried by the AGN wind particles, $\Phi_0$ is the energy flux carried by the background CRs, and $y$ and $y_0$ are the perturbed and unperturbed stratospheric NO abundances, respectively, expressed in parts per billion.

The quantity $R_{\mathrm{NO}}$ can also be expressed in the following way:

$$y = \frac{R_{\mathrm{NO}}\, t_{\mathrm{NO}}}{\sigma_{\mathrm{strat}}} \times 10^9, \tag{17}$$

where $\sigma_{\mathrm{strat}}$ is the stratospheric column density for EPs, taken to be $5 \times 10^{23}$ molecules cm$^{-2}$, the factor $10^9$ accounts for the fact that $y$ is expressed in parts per billion, and the quantity $t_{\mathrm{NO}}$ is the residence time for the NO in the stratosphere before diffusing out; in J. Ellis & D. N. Schramm (1995), it is assumed to be $t_{\mathrm{NO}} \approx 4$ yr. However, unlike in a supernova, where an impulsive injection is followed by the diffusion (or loss) of NO, the AGN case involves a continuous supply of energy that sustains NO production over the Salpeter time ($\Delta t_{\mathrm{Salp}}$), which is comparatively very long. In our model, we therefore neglect the removal of NO from the stratosphere and assume that NO production is sustained for the whole duration of the AGN phase, which we approximate to $\Delta t_{\mathrm{Salp}}$.

Solving Equation (17) for $R_{\mathrm{NO}}$ and substituting into Equation (16) gives an equation describing the NO concentrations:

$$y^2 + 10y - \frac{R_0\, \Phi}{\Phi_0} \frac{(10 + y_0)\, \Delta t_{\mathrm{Salp}} \times 10^9}{\sigma_{\mathrm{strat}}} = 0, \tag{18}$$

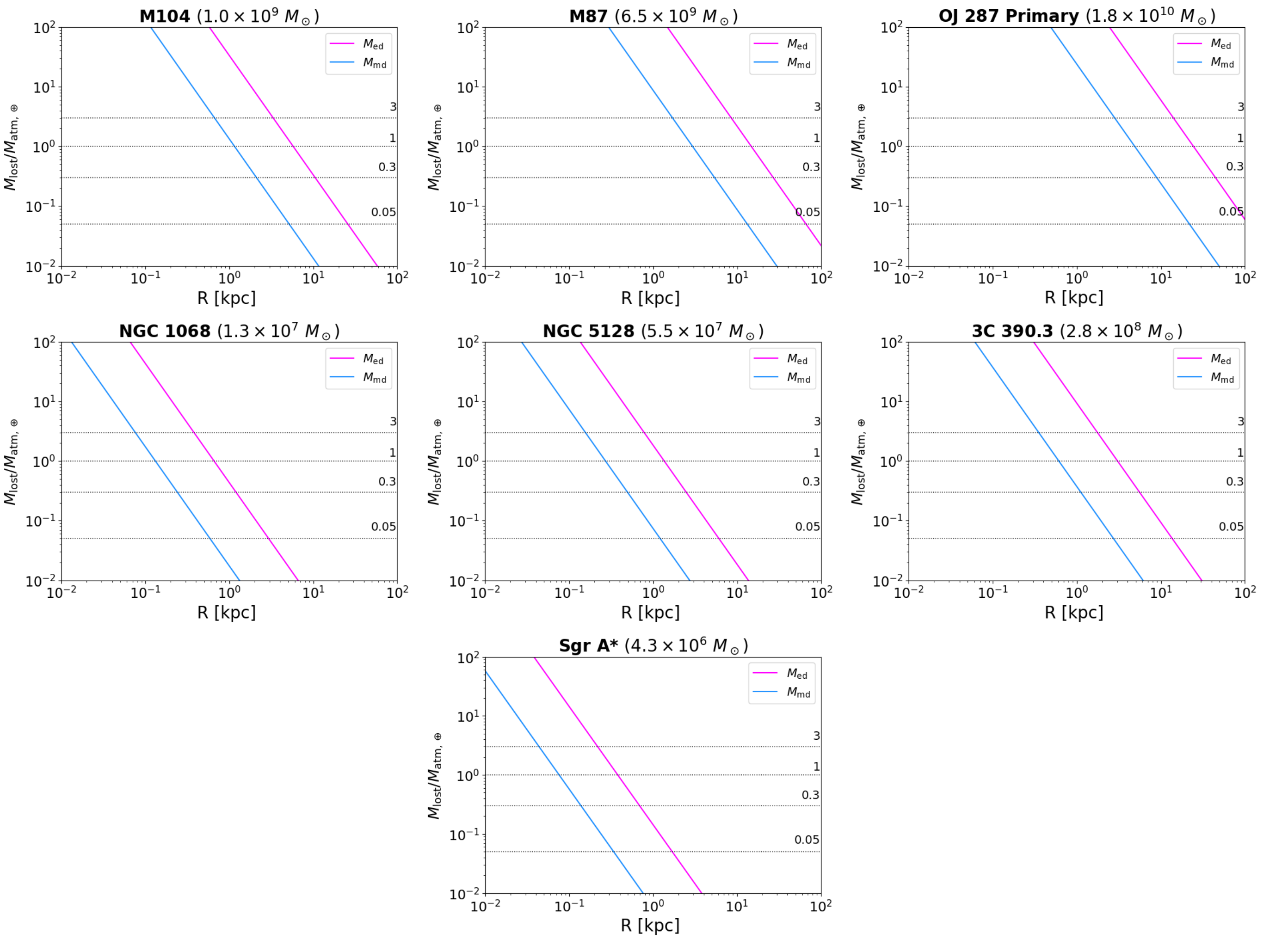

**Figure 6.** Atmospheric mass loss (relative to Earth's atmospheric mass) due to momentum- and energy-driven wind-mediated escape as a function of the distance to the central galactic SMBH (in kiloparsecs).

where $\Phi_0 \approx 9 \times 10^4\,\mathrm{erg\,cm^{-2}\,yr^{-1}}$, $R_0 \approx 9 \times 10^{14}\,\mathrm{molecules\,cm^{-2}\,yr^{-1}}$, and $y_0 = 3\,\mathrm{ppb}$ (A. Ambrifi et al. 2022).

This can be simplified to

$$y^2 + 10y - k\,\Phi\Delta t_{\mathrm{Salp}} = 0, \tag{19}$$

where $k$ is defined as

$$k = \frac{R_0}{\Phi_0}\frac{10^9(10 + y_0)}{\sigma_{\mathrm{strat}}}.$$

The solution for $y$ is then

$$y = -5 + \frac{1}{2}\sqrt{100 + 4(k\,\Phi\Delta t_{\mathrm{Salp}})}, \tag{20}$$

in which the plus sign indicates the limitation to only the physical solution.

The work done by A. Ambrifi et al. (2022) also outlines the derivation of an equation for the average energy flux attributed to AGN wind particles, which is converted to units of $\mathrm{erg\,cm^{-2}\,yr^{-1}}$ to be consistent with those of $\Phi_0$. The flux is given by

$$\Phi = \frac{\dot{\epsilon}_{\mathrm{k}}}{16\pi R^2}, \tag{21}$$

where $\dot{\epsilon}_{\mathrm{k}} \approx 0.05 L_{\mathrm{Edd}}$ and $0.001 L_{\mathrm{Edd}}$ for energy-driven winds and momentum-driven winds, respectively.

It is important to note that the ratio $F$ of stratospheric ozone abundance in the perturbed ($[\mathrm{O_3}]$) and unperturbed ($[\mathrm{O_3}]_0$) cases can be computed using the following equations derived in A. Ambrifi et al. (2022):

$$F = \frac{[\mathrm{O}_3]}{[\mathrm{O}_3]_0} = \frac{\sqrt{16 + 9X^2} - 3X}{2}, \tag{22}$$

where $X$ represents the ratio of perturbed ($[\mathrm{NO}]$) and unperturbed ($[\mathrm{NO}]_0$) NO abundances:

$$X = \frac{[\mathrm{NO}]}{[\mathrm{NO}]_0} = \frac{y_0 + y}{y_0} = \frac{3 + y}{3}, \tag{23}$$

where the last equality is obtained by substituting the value $y_0 = 3\,\mathrm{ppb}$ introduced previously.

Substituting Equation (23) into Equation (22), we find that

$$F = \frac{1}{2}\left(\sqrt{16 + (3 + y)^2} - (3 + y)\right). \tag{24}$$

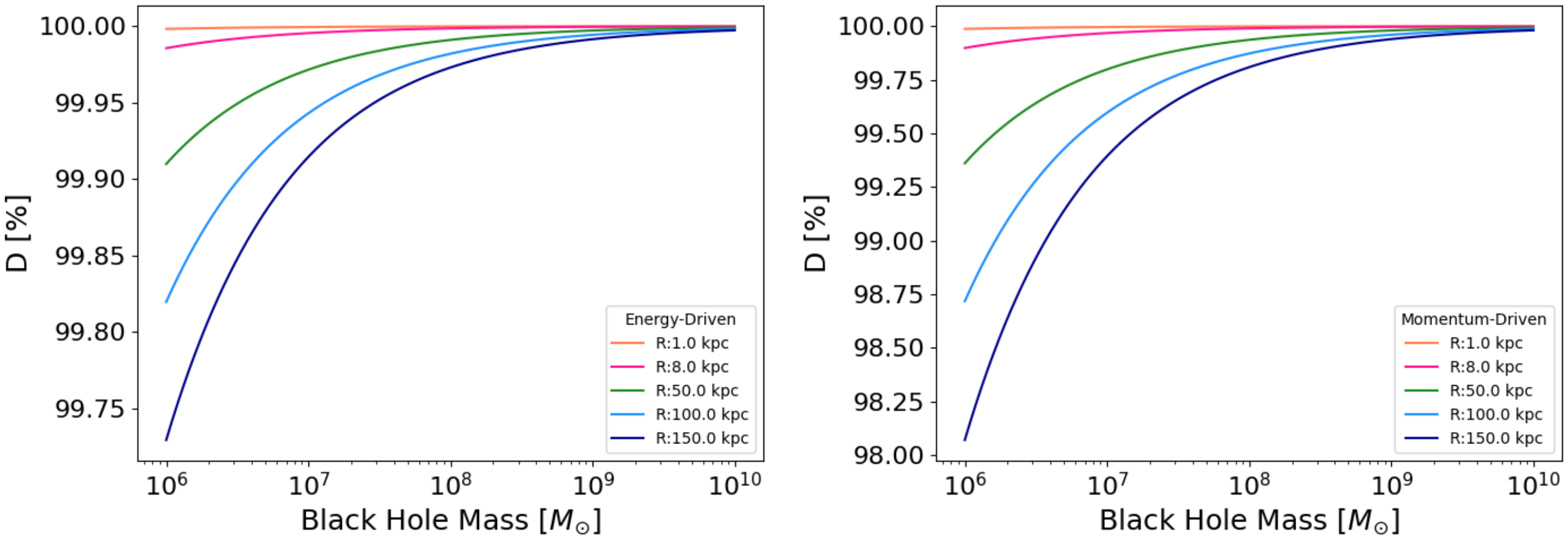

**Figure 7.** Percentage of ozone depletion in an Earth-like atmosphere (denoted by $D$) due to $NO_x$ production caused by energy- and momentum-driven AGN winds as a function of the mass of the central galactic SMBH in solar masses. The left panel shows the energy-driven case, while the right one shows the effect of momentum-driven winds. The lines represent the distance $R$ from the galactic center (in kiloparsecs).

Then, the depletion is simply given by

$$D = 1 - F = 1 - \frac{1}{2}(\sqrt{16 + (3 + y)^2} - (3 + y)). \quad (25)$$

Given their extreme velocities, it is worth examining whether AGN winds, particularly UFOs with velocities $\sim 0.1c$ and postshock speeds of $O(1000)\,\mathrm{km\,s^{-1}}$, may contribute to ozone depletion in atmospheres similar to Earth's (M. Moe et al. 2009; F. Tombesi et al. 2011, 2015; G. Vietri et al. 2018). Although high-energy particles may also facilitate the synthesis of prebiotic molecules, we do not explore this possibility due to its uncertainty and complexity (M. Lingam et al. 2018). Compared to the previous results, our focus is therefore restricted to modern Earth-like atmospheres composed of $N_2$ and $O_2$, rather than hydrogen-rich ones, to assess ozone depletion driven by nitrogen oxide ($NO_x$) production (P. J. Crutzen 1971) possibly resulting from interactions with UFOs (A. Ambrifi et al. 2022).

The percentage of ozone depletion in an Earth-like atmosphere is presented in Figures 7 and 8. For both energy-driven and momentum-driven mechanisms, the results indicate that ozone depletion increases with the mass of the BH. This trend is along expected lines because more massive BHs generate more powerful AGN winds, which lead to greater production of $NO_x$ and, consequently, more significant ozone depletion.

The figures also show that as the distance from the galactic center increases, the extent of ozone depletion decreases. The prediction matches physical intuition because the influence of AGN winds on ozone is more pronounced in regions closer to the BH and diminishes with increasing distance, which aligns with expectations. Figure 7 illustrates these general trends for both the energy- and momentum-driven cases, showing that ozone depletion weakens with increasing galactocentric distance. Figure 8 reinforces these patterns by showing ozone depletion profiles for specific galaxies, demonstrating that $\sim$100% ozone depletion occurs within 1 kpc of the galactic center and extends further out for more massive SMBHs.

Importantly, nearly all of the results shown in Figure 7 exceed 99% ozone depletion, even for relatively low-mass SMBHs. This indicates that total stratospheric ozone loss occurs throughout much of the inner galaxy, over AGN timescales of tens of thousands of years. This aligns with the result of A. Ambrifi et al. (2022) for Sgr A$^*$ and implies that the near-complete depletion of ozone may be the most universal and wide-ranging atmospheric consequence of AGN winds.

Overall, ozone depletion due to AGN winds is more significant for larger SMBHs and at smaller galactic distances. Energy-driven winds result in slightly greater depletion than momentum-driven winds. For high-mass SMBHs ($\gtrsim 10^8\,M_\odot$), ozone depletion approaches near-total levels ($\sim$100%) across galactic scales in the energy-driven case, indicating that AGN activity could render EPs uninhabitable in galaxies hosting SMBHs of this magnitude.

### *4.3. Threshold for Critical Ozone Depletion*

Ozone depletion after a general time interval can be determined using Equation (17). Instead of relying on the Salpeter timescale, we solve for a generalized time interval, $\Delta t$.

It is straightforward to express $y$ explicitly in terms of $F$, which will simplify to

$$y = \left(\frac{4}{F} - F - 3\right). \quad (26)$$

For 90% ozone depletion, $D = 1 - F = 0.9$, or $F = 0.1$. Therefore, $y = 36.9 \approx 37$ ppb. This represents a severe depletion scenario adopted to illustrate an upper-limit case. It is worth noting that even a much smaller reduction in ozone, on the order of 25%–30%, could already pose a significant threat to Earth-like biota (B. C. Thomas et al. 2005, 2018; A. L. Melott & B. C. Thomas 2011).

Then, plugging in the value for $y$ and rearranging Equation (20) for $\Delta t$ gives

$$\Delta t = \frac{1739}{k\ \Phi}, \quad (27)$$

in years.

The final set of results, Figures 9 and 10, depict the timescale over which NO is required to be active in an Earth-like atmosphere to cause 90% ozone depletion in an atmosphere. Under both circumstances, the timescale decreases linearly as the mass of the BH increases. More specifically, the timescales in the momentum-driven scenario are generally longer than those in the energy-driven case at the same BH mass and distance from the galactic center. In general, this implies that more massive BHs lead to shorter timescales required for 90% ozone depletion.

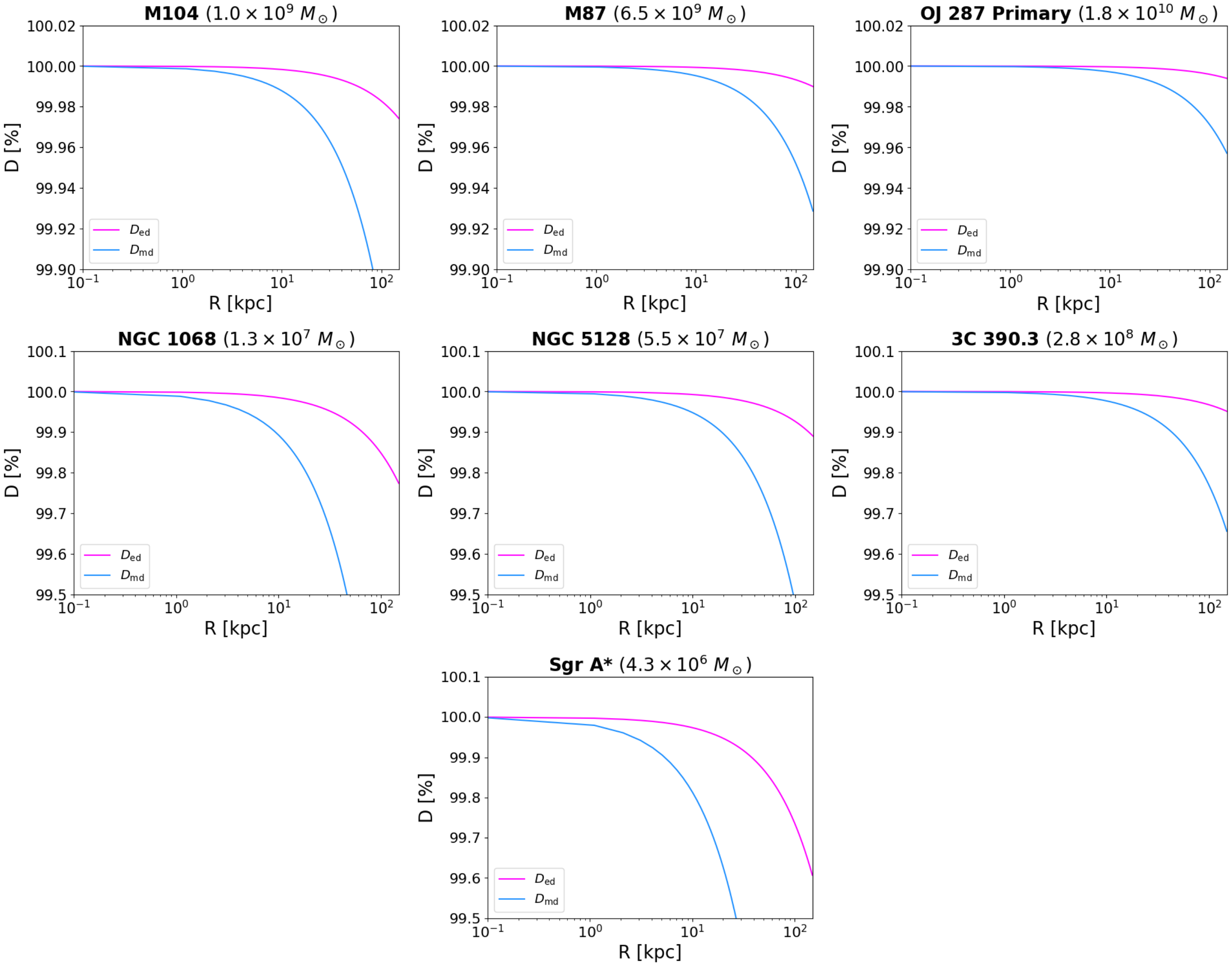

**Figure 8.** Percentage of ozone depletion in an Earth-like atmosphere (denoted by $D$) due to $NO_x$ production caused by energy- and momentum-driven AGN winds as a function of the distance to the central galactic SMBH (in kiloparsecs).

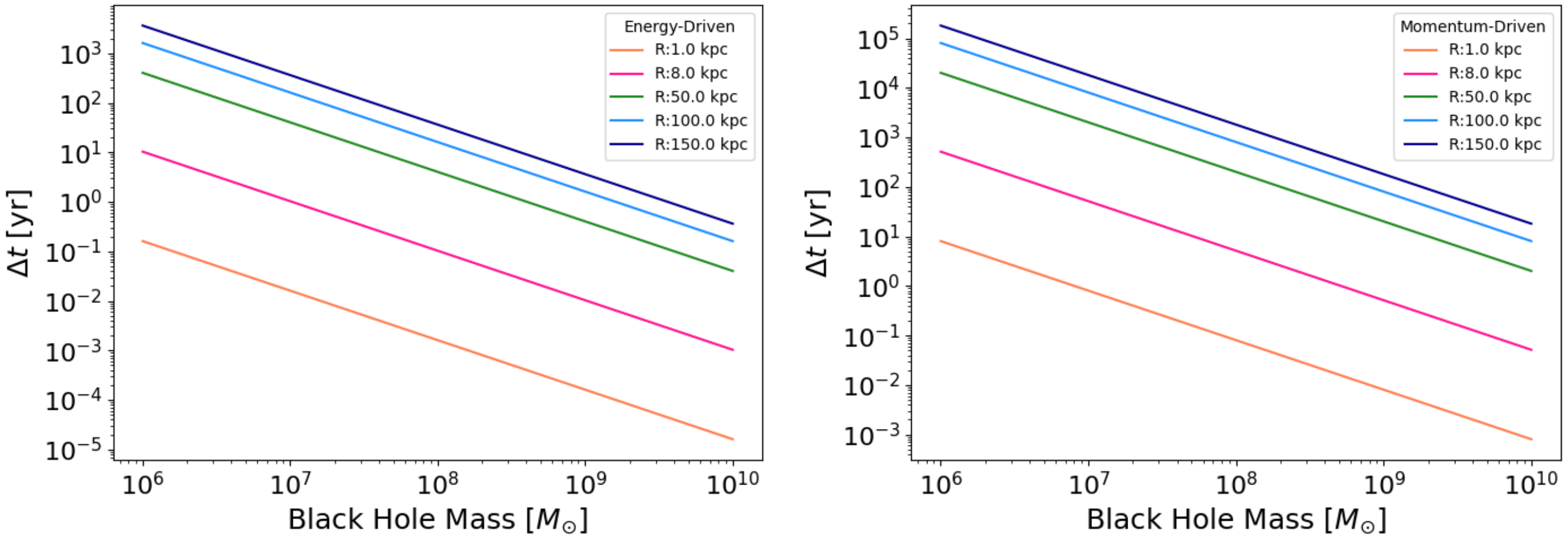

**Figure 9.** Timescale (in years) over which NO must remain consistently active in an Earth-like atmosphere to cause a 90% depletion of ozone resulting from the energy- and momentum-driven AGN wind scenarios as a function of the mass of the central galactic SMBH in solar masses. The left panel shows the energy-driven case, while the right one shows the effect of momentum-driven winds. The lines represent the distance $R$ from the galactic center (in kiloparsecs).

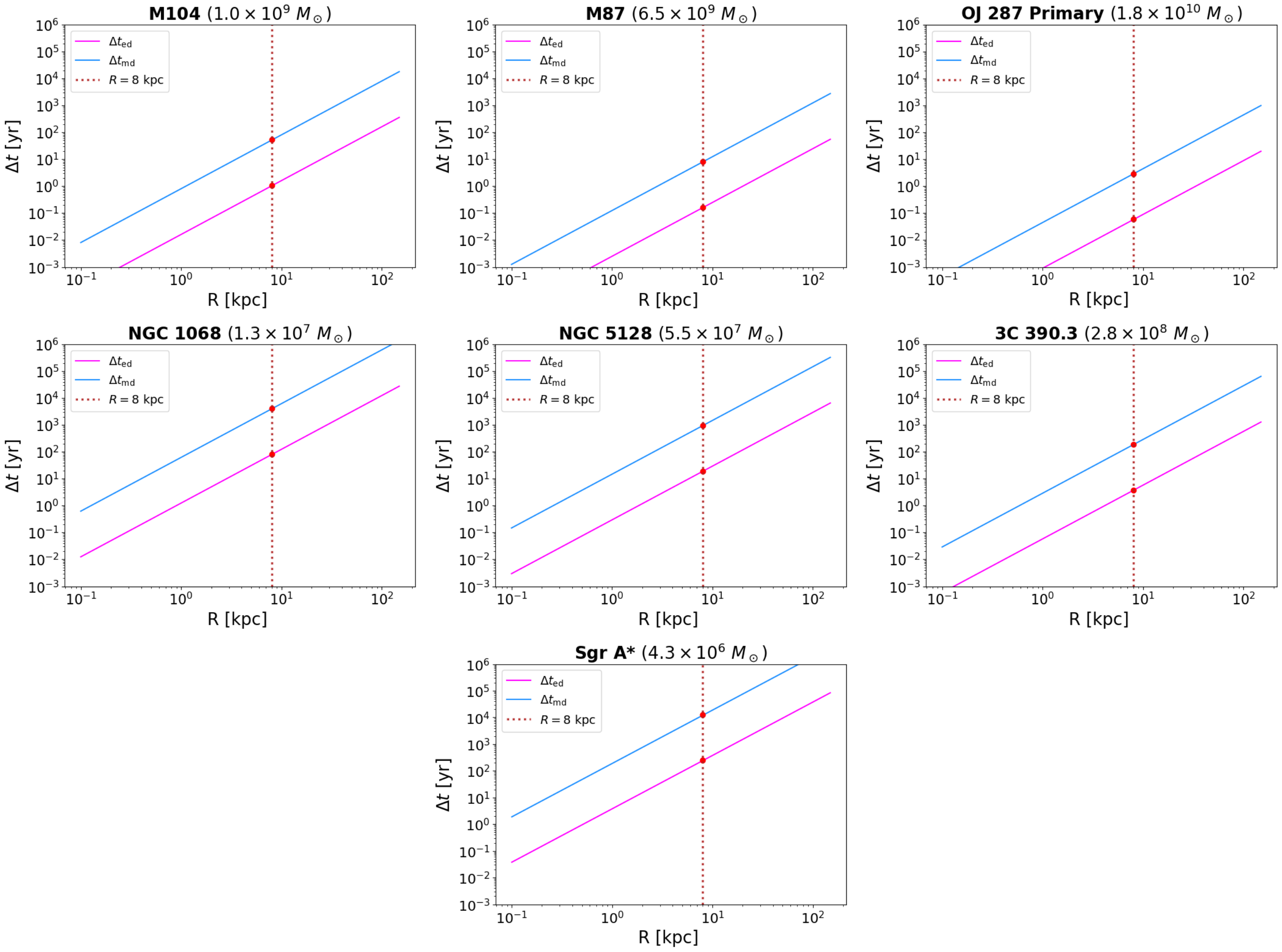

**Figure 10.** Timescale (in years) over which NO must remain consistently active in an Earth-like atmosphere to cause a 90% depletion of ozone resulting from the energy- and momentum-driven AGN wind scenarios as a function of the distance to the central galactic SMBH (in kiloparsecs). The vertical line denotes the Earth's distance from the Milky Way's center (∼8 kpc).

As stated in Section 4.2, a residence time of 4 yr is specified for NO in the stratosphere (J. Ellis & D. N. Schramm 1995). This timescale serves as a benchmark for evaluating whether AGN winds can cause significant ozone depletion before NO naturally dissipates. In effect, there is a competition of timescales, with the question becoming whether the UFO will deplete 90% of the ozone layer faster than NO is removed from the stratosphere. The aforementioned residence timescale provides a basis for estimating the distances at which ozone depletion may become significant.

## 5. Discussion and Conclusions

While the significance of SMBH activity in shaping habitability has gained popularity in recent years, the specific impact of AGN winds and outflows on planetary atmospheres remains largely unexplored in contemporary research. Therefore, this study aimed to address the relationship between the mass of the SMBH at the center of galaxies and exoplanetary habitability, in light of the substantial variation in the observed masses of SMBHs across various galaxies.

In Section 2, we improved upon previous studies and outlined a framework for investigating the atmospheric consequences of AGN winds. We introduced our model and parameter choices used to evaluate the influence of both energy- and momentum-driven UFOs on planetary atmospheres across a range of galactic distances and SMBH masses. Importantly, our analysis is restricted to planets located within the region of the wind, and does not consider regions outside of it. This section also detailed the theoretical assumptions used to assess atmospheric heating and escape.

The results of this study are presented and discussed in Section 3 and Section 4. We found that energy-driven UFOs lead to significantly higher atmospheric heating than momentum-driven winds, particularly at small galactocentric radii. The heating effect drops off rapidly with distance and is subdominant for momentum-driven cases. The most probable molecular velocities in the planetary atmosphere exceed the escape velocity of the Earth within the inner few kiloparsecs of galaxies hosting SMBHs with masses $\geqslant 10^7\,M_{\odot}$, as seen from the left panel in Figure 3, where the energy-driven winds achieve this threshold more efficiently than the momentum-driven winds. In contrast, momentum-driven winds require a minimum SMBH mass $\geqslant 6 \times 10^8 M_\odot$, shown in the right panel of Figure 3, for molecular velocities to exceed Earth's escape velocity.

As we noted in Section 3.2, our results show that there is a regime where the most probable velocity $v_{\rm mp}$ surpasses the escape velocity $v_{\rm esc}$, which implies catastrophic atmospheric loss and ionization. We further showed that the fraction of the atmosphere lost increases with BH mass in both scenarios, with momentum-driven winds consistently yielding lower values. A somewhat more massive planet, with a higher escape velocity $v_{\rm esc}$, would change the magnitude of this effect commensurately, but the same overall trends would be seen. The emphasis on modeling atmospheric escape in Section 3 is because the complete loss of an atmosphere would, according to the phase diagram of water, negate the existence of liquid water on the surface due to the absence of surface pressure exerted by the atmosphere. It would also lead to elevated fluxes of UV radiation and CRs reaching the surface. Each of these effects are detrimental to planetary habitability for biological organisms on the surface (M. Lingam & A. Loeb 2021), especially the inability to host liquid water, which is considered one of the requirements for life as we know it.

We also explored ozone depletion in Earth-like atmospheres, showing that the extent of depletion increases with BH mass and proximity to the AGN. This is due to more massive BHs producing stronger winds and more $NO_x$, which accelerate ozone loss. The decreasing ozone depletion with distance from the galactic center validates that AGN wind effects are strongest near the BH. Once again, energy-driven winds cause slightly more depletion than momentum-driven ones. For SMBHs with masses $\geqslant 10^8 M_\odot$, ozone loss approaches 100% in the energy-driven case and remains $\geqslant$99.75% in the momentum-driven case across galactic scales, as shown in Figure 7.

These simulations suggest that, for the most massive SMBHs, the effective region of influence extends well beyond the inner galaxy and potentially includes the galactic halo in the energy-driven scenario. However, this extent could be mitigated by a sufficiently dense ISM, particularly in environments rich in molecular gas and dust. It should be noted that our model does not currently account for the effects of absorption or scattering by material in the ISM, which might however be more significant for electromagnetic radiation than high-energy particles, and therefore likely represents an upper limit on the true extent of atmospheric impact.

Using a 4 yr NO residence time in the stratosphere, we can estimate the distances from the galactic center where AGN-driven ozone depletion becomes significant. As SMBH mass increases, this critical distance also increases for both wind types. At Earth's distance from the Milky Way's center ($\sim$8 kpc), the timescale for 90% ozone loss can be estimated across different SMBH masses, highlighting the potential risk of AGN activity even at large galactic distances. This timescale, depicted in Figure 9, decreases linearly with increasing BH mass, indicating that more massive BHs cause faster ozone loss. In general, momentum-driven cases require longer timescales than energy-driven ones under the same conditions.

The timescale in which 90% of the ozone layer is depleted represents an important benchmark to determining when potential biological impacts may occur during the "lifespan" of an exoplanet, though it must be understood that organisms in high-radiation environments might have evolved some degree of evolutionary adaptations (M. Lingam & A. Balbi 2024). It is possible that even after only 30% of the ozone is depleted, the amount of UV-B radiation from the host star that reaches the surface doubles and can potentially trigger mass extinctions (N. Gehrels et al. 2003; B. C. Thomas et al. 2005; A. L. Melott & B. C. Thomas 2011). As a consequence, it is anticipated that a 90% depletion would lead to a severalfold increase in UV-B exposure, likely causing severe ecological harm (A. Segura et al. 2010; M. Lingam & A. Loeb 2017). However, this outcome does not guarantee the extinction of all life, as subsurface life may remain viable (C. Magnabosco et al. 2018; M. Lingam & A. Loeb 2019b), as well as organisms underwater or in protected niches (C. S. Cockell 2000).

Results derived from Figure 10 are shown in Table 1, specifically the distance from the galactic center where UFOs from AGN increase with the mass of the SMBH for both the energy- and momentum-driven cases. These distance values are found by setting $\Delta t \approx 4$ yr, previously mentioned as the residence timescale for NO in the stratosphere. The results are as expected; intuitively, a more massive BH should influence regions at greater distances via its UFOs.

Quantitatively, the characteristic distances listed in Table 1 follow an approximately $R \propto M_{\rm BH}^{1/2}$ scaling, with fitted exponents ranging from 0.49 to 0.62 across the different regimes. This scaling can be derived analytically using Equations (1) and (4), where $L_{\rm Edd} \propto M_{\rm BH}$ and $\dot{\epsilon} \propto L_{\rm Edd}$. From these relations, both Equations (21) and (27) yield $R \propto M_{\rm BH}^{1/2}$. This near-square-root dependence indicates that the extent of AGN-driven atmospheric influence primarily reflects flux conservation: the distance at which each effect occurs increases roughly with the square root of the SMBH mass, as expected if the flux (or deposited kinetic energy per unit area) remains constant.

The mass-dependent trends identified are consistent with recent observational findings that link UFO properties directly to SMBH mass. In particular, large-sample analyses have shown that both the velocity and kinetic power of UFOs increase with BH mass, as implied by Equation (4) (see, e.g., Figure 7 of V. E. Gianolli et al. 2024).

Our results build upon and significantly expand those published in the precursor study by A. Ambrifi et al. (2022), which accounted only for the effects of Sgr A* on the Milky Way. By extending the BH mass range to $10^{10} M_\odot$, our work captures more extreme AGN-driven impacts on planet atmospheres, revealing the magnitudes of mass loss and ozone depletion, which exceed those found in the previous study.

A recent study by K. I. Sippy et al. (2025) similarly investigates the influence of AGN radiation on planetary atmospheres, with a particular emphasis on surface UV flux and its implications for habitability. Their simulations, using the PALEO photochemical-climate model, suggest that any significant UV impact is confined to the innermost regions of galaxies like the Milky Way and M87, primarily affecting the Galactic bulge (within $\leqslant$1 kpc of the center), even under assumptions of no attenuation by the ISM. Notably, the K. I. Sippy et al. (2025) model is dedicated to analyzing only the impact of electromagnetic radiation, whereas our model considers an AGN high-energy particle wind characterized by a UFO operating at its maximum kinetic power.

Since UV radiation from AGN initiates photochemical reactions that modify atmospheric composition, models like PALEO are particularly well suited to capture these consequences. PALEO's ability to simulate both surface conditions and atmospheric chemistry in exoplanetary environments permits a more comprehensive investigation of how AGN influence atmospheric evolution and habitability. In future work, the deployment of similar models may enable an improved understanding of the consequences of the $NO_x$ species in the presence of an increased CR flux, such as those arising from AGN winds or supernovae.

The primacy of high-energy particles in mediating atmospheric consequences is well established and not exclusive to AGN. As shown by A. L. Melott et al. (2017), in the case of a supernova at 50 pc, the accompanying UV radiation is still less intense than solar UV radiation; yet the ozone layer is depleted by approximately 60%, primarily due to the impact of energetic particles. This particle-driven mechanism is analogous to the AGN scenario studied throughout this paper, where it is not the intrinsic UV flux from the AGN that proves to be most lethal, but rather the increase in surface UV radiation resulting from particle-induced chemical alterations that deteriorate the protective ozone layer. Consequently, the AGN "kill zone" should be understood as a particle-mediated phenomenon.

While prior studies such as W. Ishibashi (2024) highlight the vulnerability of exoplanets in the Milky Way's Galactic bulge due to XUV-driven atmospheric photoevaporation, our results suggest that AGN winds may influence planetary environments at much larger galactocentric radii than UV or XUV radiation alone. This implies that kinetic feedback from AGN activity could extend the zone of impact well beyond radiation-based kill zones. Since our current model does not incorporate radiative effects, the combined influence of winds and high-energy radiation on the Galactic habitable zone (C. H. Lineweaver et al. 2004; N. Prantzos 2008) should be explored in future studies.

Having delved into the ramifications of varying masses of SMBHs, the subsequent phase of research involves the introduction of variable AGN wind speeds and incorporates interstellar gas interactions to create a more realistic model of AGN-driven atmospheric impacts. This includes examining how mass-loaded winds absorb radiation and alter CR flux, as well as assessing the influence of shocks and intervening matter on the UV and X-ray radiation received by a planet. Incorporating atmospheric chemistry via a model like PALEO would enable a more thorough investigation of AGN-induced atmospheric changes, extending the analysis to include effects such as CR ionization and atmospheric ablation in addition to ozone depletion. While our present work only considered molecular-hydrogen- and nitrogen-dominated atmospheres, future efforts could explore a broader range of atmospheric compositions, such as those analogous to Venus or Titan, to better capture the diversity of exoplanetary environments.

## Acknowledgments

J.W. thanks E.P. for continuous guidance, our collaborators for valuable feedback, and her dog, Kody, for unwavering emotional support throughout this work. A.A. acknowledges support by the Spanish Agencia estatal de investigación via PID2021-124879NB-I00.

## ORCID iDs

Jourdan Waas https://orcid.org/0009-0008-3084-3769
Eric S. Perlman https://orcid.org/0000-0002-3099-1664
Manasvi Lingam https://orcid.org/0000-0002-2685-9417
Emily Lohmann https://orcid.org/0009-0006-9780-8445
Jackson Kernan https://orcid.org/0009-0002-1291-9279
Francesco Tombesi https://orcid.org/0000-0002-6562-8654
Amedeo Balbi https://orcid.org/0000-0002-3929-6932
Alessandra Ambrifi https://orcid.org/0000-0003-1536-4213

## References

Airapetian, V. S., Barnes, R., Cohen, O., et al. 2020, IJAsB, 19, 136
Aird, J., Coil, A. L., & Georgakakis, A. 2018, MNRAS, 474, 1225
Akiyama, K., Alberdi, A., Alef, W., et al. 2019, ApJL, 875, L6
Amaro-Seoane, P., & Chen, X. 2019, JCAP, 2019, 056
Ambrifi, A., Balbi, A., Lingam, M., Tombesi, F., & Perlman, E. 2022, MNRAS, 512, 505
Auchettl, K., Ramirez-Ruiz, E., & Guillochon, J. 2018, ApJ, 852, 37
Balbi, A., & Tombesi, F. 2017, NatSR, 7, 16626
Beech, M. 2011, Ap&SS, 336, 287
Bischetti, M., Piconcelli, E., Vietri, G., et al. 2017, A&A, 598, A122
Boehle, A., Ghez, A. M., Schödel, R., et al. 2016, ApJ, 830, 17
Brain, D. A., Bagenal, F., Ma, Y. J., Nilsson, H., & Stenberg Wieser, G. 2016, JGRE, 121, 2364
Brunton, I. R., O'Mahoney, C., Fields, B. D., Melott, A. L., & Thomas, B. C. 2023, ApJ, 947, 42
Canfield, D. E., Glazer, A. N., & Falkowski, P. G. 2010, Sci, 330, 192
Cappellari, M., Neumayer, N., Reunanen, J., et al. 2009, MNRAS, 394, 660
Catling, D. C., & Kasting, J. F. 2017, Atmospheric Evolution on Inhabited and Lifeless Worlds (Cambridge Univ. Press)
Chartas, G., Cappi, M., Vignali, C., et al. 2021, ApJ, 920, 24
Chen, H., Forbes, J. C., & Loeb, A. 2018, ApJL, 855, L1
Cockell, C. S. 2000, OLEB, 30, 467
Congiu, E., Contini, M., Ciroi, S., et al. 2017, FrASS, 4, 27
Cramer, E. S., Briggs, M. S., Liu, N., et al. 2017, GeoRL, 44, 5240
Crnojević, D., Sand, D. J., Spekkens, K., et al. 2016, ApJ, 823, 19
Crutzen, P. J. 1971, JGR, 76, 7311
Crutzen, P. J., Isaksen, I. S. A., & Reid, G. C. 1975, Sci, 189, 457
Czerny, B., Witt, H. J., & Zycki, P. 1997, ESA Special Publication, 382, 397
Dartnell, L. R. 2011, AsBio, 11, 551
Debuhr, J., Quataert, E., & Ma, C.-P. 2012, MNRAS, 420, 2221
Doherty, M., Arnaboldi, M., Das, P., et al. 2009, A&A, 502, 771
Dolgov, A. D. 2020, JPhCS, 1690, 012183
Dong, C., Jin, M., Lingam, M., et al. 2018a, PNAS, 115, 260
Dong, C., Lingam, M., Ma, Y., & Cohen, O. 2017, ApJL, 837, L26
Dong, C., Lee, Y., Ma, Y., et al. 2018b, ApJL, 859, L14
Elkins-Tanton, L. T., & Seager, S. 2008, ApJ, 685, 1237
Ellis, J., & Schramm, D. N. 1995, PNAS, 92, 235
Event Horizon Telescope Collaboration, Akiyama, K., Alberdi, A., et al. 2022, ApJL, 930, L12
Faucher-Giguère, C.-A., & Quataert, E. 2012, MNRAS, 425, 605
Forbes, J. C., & Loeb, A. 2018, MNRAS, 479, 171
Gebhardt, K., Bender, R., Bower, G., et al. 2000, ApJL, 539, L13
Gehrels, N., Laird, C. M., Jackman, C. H., et al. 2003, ApJ, 585, 1169
Ghez, A. M., Salim, S., Weinberg, N. N., et al. 2008, ApJ, 689, 1044
Gianolli, V. E., Bianchi, S., Petrucci, P.-O., et al. 2024, A&A, 687, A235
Gillessen, S., Eisenhauer, F., Trippe, S., et al. 2009, ApJ, 692, 1075
Gofford, J., Reeves, J. N., Tombesi, F., et al. 2013, MNRAS, 430, 60
Gronoff, G., Arras, P., Baraka, S., et al. 2020, JGRA, 125, e27639
Hainline, K. N., Hickox, R. C., Chen, C.-T., et al. 2016, ApJ, 823, 42
Hanslmeier, A. 2017, Supernovae, Our Solar System, and Life on Earth (Springer), 2489
Heinz, S. 2022, MNRAS, 513, 4669
Hewett, P. C., & Foltz, C. B. 2003, AJ, 125, 1784
Igo, Z., Parker, M. L., Matzeu, G. A., et al. 2020, MNRAS, 493, 1088
Ishibashi, W. 2024, MNRAS, 533, 455
Kennard, E. H. E. H. 1938, Kinetic Theory of Gases: With an Introduction to Statistical Mechanics (first ed.; McGraw-Hill)
King, A., & Pounds, K. 2015, ARA&A, 53, 115
Kormendy, J., Bender, R., Ajhar, E. A., et al. 1996, ApJ, 473, L91
Kubyshkina, D., Way, M. J., Dandouras, I., et al. 2026, SSRv, 222, 28
Laha, S., Reynolds, C. S., Reeves, J., et al. 2021, NatAs, 5, 13

Lillis, R. J., Deighan, J., Fox, J. L., et al. 2017, JGRA, 122, 3815
Lineweaver, C. H. 2025, Galactic Habitability, Vol. 305 (Oxford Univ. Press)
Lineweaver, C. H., Fenner, Y., & Gibson, B. K. 2004, Sci, 303, 59
Lingam, M. 2019, ApJL, 874, L28
Lingam, M., & Balbi, A. 2024, From Stars to Life: A Quantitative Approach to Astrobiology (Cambridge Univ. Press)
Lingam, M., Dong, C., Fang, X., Jakosky, B. M., & Loeb, A. 2018, ApJ, 853, 10
Lingam, M., Ginsburg, I., & Bialy, S. 2019, ApJ, 877, 62
Lingam, M., & Loeb, A. 2017, ApJ, 848, 41
Lingam, M., & Loeb, A. 2018, JCAP, 2018, 020
Lingam, M., & Loeb, A. 2019a, RvMP, 91, 021002
Lingam, M., & Loeb, A. 2019b, IJAsB, 18, 112
Lingam, M., & Loeb, A. 2020, ApJL, 901, L11
Lingam, M., & Loeb, A. 2021, Life in the Cosmos: From Biosignatures to Technosignatures (Harvard Univ. Press)
Lister, M. L., Homan, D. C., Kellermann, K. I., et al. 2021, ApJ, 923, 30
Liu, C., Chen, X., & Du, F. 2020, ApJ, 899, 92
Longobardi, A., Arnaboldi, M., Gerhard, O., & Hanuschik, R. 2015, A&A, 579, A135
Luminari, A., Nicastro, F., Elvis, M., et al. 2021, A&A, 646, A111
Madhusudhan, N., Piette, A. A. A., & Constantinou, S. 2021, ApJ, 918, 1
Magnabosco, C., Lin, L. H., Dong, H., et al. 2018, NatGe, 11, 707
Marconi, A., Risaliti, G., Gilli, R., et al. 2004, MNRAS, 351, 169
Martini, P., & Weinberg, D. H. 2001, ApJ, 547, 12
McMahon, S., O'Malley-James, J., & Parnell, J. 2013, P&SS, 85, 312
Melott, A. L., & Thomas, B. C. 2011, AsBio, 11, 343
Melott, A. L., Thomas, B. C., Kachelrieß, M., Semikoz, D. V., & Overholt, A. C. 2017, ApJ, 840, 105
Mestici, S., Tombesi, F., Gaspari, M., Piconcelli, E., & Panessa, F. 2024, MNRAS, 532, 3036
Mizumoto, M., Izumi, T., & Kohno, K. 2019, ApJ, 871, 156
Moe, M., Arav, N., Bautista, M. A., & Korista, K. T. 2009, ApJ, 706, 525
Owen, J. E. 2019, AREPS, 47, 67
Pacetti, E., Balbi, A., Lingam, M., Tombesi, F., & Perlman, E. 2020, MNRAS, 498, 3153
Peterson, W. K., Brain, D. A., Schnepf, N. R., et al. 2024, GeoRL, 51, e2023GL107675
Pounds, K., Lobban, A., Reeves, J., & Vaughan, S. 2016, MNRAS, 457, 2951
Pounds, K. A., King, A. R., Page, K. L., & O'Brien, P. T. 2003, MNRAS, 346, 1025
Prantzos, N. 2008, SSRv, 135, 313
Rankine, A. L., Hewett, P. C., Banerji, M., & Richards, G. T. 2020, MNRAS, 492, 4553
Scherf, M., Lammer, H., & Spross, L. 2024, AsBio, 24, e916
Schwartz, S. E. 2007, JGRD, 112, D24S05
Seager, S., Bains, W., & Hu, R. 2013, ApJ, 777, 95
Seager, S., Huang, J., Petkowski, J. J., & Pajusalu, M. 2020, NatAs, 4, 802
Segura, A., Walkowicz, L. M., Meadows, V., Kasting, J., & Hawley, S. 2010, AsBio, 10, 751
Sergeev, S. G., Nazarov, S. V., & Borman, G. A. 2016, MNRAS, 465, 1898
Shen, Y. 2013, BASI, 41, 61
Sippy, K. I., Eager-Nash, J. K., Hickox, R. C., Mayne, N. J., & Brumback, M. C. 2025, ApJ, 980, 221
Steinhardt, C. L., & Elvis, M. 2010, MNRAS, 402, 2637
Sun, A.-L., Greene, J. E., & Zakamska, N. L. 2017, ApJ, 835, 222
Thomas, B. C. 2018, AsBio, 18, 481
Thomas, B. C., & Yelland, A. M. 2023, ApJ, 950, 41
Thomas, B. C., Melott, A. L., Jackman, C. H., et al. 2005, ApJ, 634, 509
Tombesi, F., Cappi, M., Reeves, J. N., et al. 2013, MNRAS, 430, 1102
Tombesi, F., Cappi, M., Reeves, J. N., et al. 2011, ApJ, 742, 44
Tombesi, F., Cappi, M., Reeves, J. N., et al. 2010a, A&A, 521, A57
Tombesi, F., Meléndez, M., Veilleux, S., et al. 2015, Natur, 519, 436
Tombesi, F., Sambruna, R. M., Reeves, J. N., et al. 2010b, ApJ, 719, 700
Tombesi, F., Tazaki, F., Mushotzky, R. F., et al. 2014, MNRAS, 443, 2154
Valtonen, M. J., Zola, S., Ciprini, S., et al. 2016, ApJL, 819, L37
Vestergaard, M., & Gültekin, K. 2023, The Encyclopedia of Cosmology, 3 (World Scientific) 227–304
Vietri, G., Piconcelli, E., Bischetti, M., et al. 2018, A&A, 617, A81
Voigt, C., Marushchak, M. E., Lamprecht, R. E., et al. 2017, PNAS, 114, 6238
Wang, J.-M., Songsheng, Y.-Y., Li, Y.-R., Du, P., & Yu, Z. 2020, MNRAS, 497, 1020
Weymann, R. J., Morris, S. L., Foltz, C. B., & Hewett, P. C. 1991, ApJ, 373, 23
Whitman, W. B., Coleman, D. C., & Wiebe, W. J. 1998, PNAS, 95, 6578
Wisłocka, A. M., Kovačević, A. B., & Balbi, A. 2019, A&A, 624, A71
Wyithe, J. S. B., & Loeb, A. 2003, ApJ, 595, 614
Xu, X., Arav, N., Miller, T., & Benn, C. 2019, ApJ, 876, 105
Xu, Y., Pinto, C., Rogantini, D., et al. 2024, A&A, 687, A179
Zhu, W., & Dong, S. 2021, ARA&A, 59, 291
Zubovas, K. 2018, MNRAS, 473, 3525
Zubovas, K., Bialopetravičius, J., & Kazlauskaitė, M. 2022, MNRAS, 515, 1705
Zubovas, K., & King, A. 2012, ApJL, 745, L34